# Small Approximate Pareto Sets for Bi-objective Shortest Paths and Other Problems[*]

Ilias Diakonikolas[†]
ilias@cs.columbia.edu

Mihalis Yannakakis[‡]
mihalis@cs.columbia.edu

## Abstract

We investigate the problem of computing a minimum set of solutions that approximates within a specified accuracy $\epsilon$ the Pareto curve of a multiobjective optimization problem. We show that for a broad class of bi-objective problems (containing many important widely studied problems such as shortest paths, spanning tree, and many others), we can compute in polynomial time an $\epsilon$-Pareto set that contains at most twice as many solutions as the minimum such set. Furthermore we show that the factor of 2 is tight for these problems, i.e., it is NP-hard to do better. We present upper and lower bounds for three or more objectives, as well as for the dual problem of computing a specified number $k$ of solutions which provide a good approximation to the Pareto curve.

---

[*]A preliminary version of this work appeared in the Proceedings of the 10th International Workshop on Approximation Algorithms for Combinatorial Optimization Problems (*APPROX' 07*), 2007, pp. 74–88.

[†]Supported by NSF grant CCF-04-30946, NSF grant CCF-07-28736 and an Alexander S. Onassis Foundation Fellowship.

[‡]Supported by NSF grant CCF-04-30946 and NSF grant CCF-07-28736.

# 1 Introduction

In many decision making situations it is typically the case that more than one criteria come into play. For example, when purchasing a product (car, tv, etc.) we care about its cost, quality, etc. When choosing a route we may care about the time it takes, the distance travelled, etc. When designing a network we may care about its cost, its capacity (the load it can carry), its coverage. This type of *multicriteria* or *multiobjective* problems arise across many diverse disciplines, in engineering, in economics and business, healthcare, and others. The area of multiobjective optimization has been (and continues to be) extensively investigated in the management science and optimization communities with many papers, conferences and books (see e.g. [Cli, Ehr, EG, FGE, Mit]).

In multiobjective problems there is typically no uniformly best solution in all objectives, but rather a trade-off between the different objectives. This is captured by the *trade-off* or *Pareto curve*, the set of all solutions whose vector of objective values is not dominated by any other solution. The trade-off curve represents the range of reasonable "optimal" choices in the design space; they are precisely the optimal solutions for all possible global "utility" functions that depend monotonically on the different objectives. A decision maker, presented with the trade-off curve, can select a solution that corresponds best to his/her preferences; of course different users generally may have different preferences and select different solutions. The problem is that the trade-off curve has typically exponential size (for discrete problems) or is infinite (for continuous problems), and hence we cannot construct the full curve. Thus, we have to contend with an approximation of the curve: We want to compute efficiently and present to the decision makers a small set of solutions (as small as possible) that represents as well as possible the whole range of choices, i.e. that provides a good approximation to the Pareto curve. Indeed this is the underlying goal in much of the research in the multiobjective area, with many heuristics proposed, usually however without any performance guarantees or complexity analysis as we do in theoretical computer science.

In recent years we initiated a systematic investigation [PY1, VY] to develop the theory of multiobjective approximation along similar rigorous lines as the approximation of single objective problems. The approximation to the Pareto curve is captured by the concept of an $\epsilon$-*Pareto set*, a set $P_\epsilon$ of solutions that approximately dominates every other solution; that is, for every solution $s$, the set $P_\epsilon$ contains a solution $s'$ that is within a factor $1 + \epsilon$ of $s$, or better, in all the objectives. (As usual in approximation, it is assumed that all objective functions take positive values.) Such an approximation was studied before for certain problems, e.g. multiobjective shortest paths, for which Hansen [Han] and Warburton [Wa] showed how to construct an $\epsilon$-Pareto set in polynomial time (for fixed number of objectives). Note that typically in most real-life multiobjective problems the number of objectives is small. In fact, the great majority of the multiobjective literature concerns the case of two objectives.

Consider a multiobjective problem with $d$ objectives, for example shortest path with cost and time objectives. For a given instance, and error tolerance $\epsilon$, we would like to compute a smallest set of solutions that form an $\epsilon$-Pareto set. Can we do it in polynomial time? If not, how well can we approximate the smallest $\epsilon$-Pareto set? Note that an $\epsilon$-Pareto set is not unique: in general there are many such sets, some of which can be very small and some very large. First, to have any hope we must ensure that there exists at least a polynomial size $\epsilon$-Pareto set. Indeed, in [PY1] it was shown that this is the case for every multiobjective problem with a fixed number of polynomially computable objectives. Second we must be able to construct at least one such set in polynomial time. This is not always possible. A necessary and sufficient condition for polynomial computability for all $\epsilon > 0$ is the existence of a polynomial algorithm for the following *Gap problem*: Given a vector of values $b$, either compute a solution that dominates $b$, or determine that no solution dominates $b$ by at least a factor $1 + \epsilon$ (in all the objectives). Many multiobjective problems were shown to have such a routine for the Gap problem (and many others have been shown subsequently).

Construction of a polynomial-size approximate Pareto set is useful, but not good enough in itself: For example, if we plan a trip, we want to examine just a few possible routes, not a polynomial number in the size of the map. More generally, in typical multicriteria situations, the selected representative solutions are investigated more thoroughly by the decision maker (designer, physician, corporation, etc.) to assess the different choices and pick the most preferable one, based possibly on additional factors that are perhaps not formalized or not even quantifiable. We thus want to select as small a set as possible that achieves a desired approximation. In [VY] the problem of constructing a minimum $\epsilon$-Pareto set was raised formally and investigated in a general framework. It was shown that for all bi-objective problems with a polynomial-time Gap routine, one can construct an $\epsilon$-Pareto set that contains at most 3 times the number of points of the smallest such set; furthermore, the factor 3 is best possible in the sense that for some problems it is NP-hard to do better. Further results were shown for 3 and more objectives, and for other related questions. Note that although the factor 3 of [VY] is best possible in general for two objectives, one may be able to do better for specific problems.

We show in this paper, that for an important class of bi-objective problems (containing many widely studied natural ones such as shortest paths, spanning tree, knapsack, scheduling problems and others) we can obtain a 2-approximation, and furthermore the factor of 2 is tight for them, i.e., it is NP-hard to do better. Our algorithm is a general algorithm that relies on a routine for a stronger version of the Gap problem, namely a routine that solves approximately the following *Restricted problem*: Given a (hard) bound $b_1$ for one objective, compute a solution that optimizes approximately the second objective subject to the bound. Many problems (e.g. shortest paths, etc.) have a polynomial time approximation scheme for the Restricted problem. For all such problems, a 2-approximation to the minimum $\epsilon$-Pareto set can be computed in polynomial time. Furthermore, the number of calls to the Restricted routine (and an associated equivalent dual routine) is linear in the size $\mathrm{OPT}_\epsilon$ of the optimal $\epsilon$-Pareto set.

The bi-objective shortest path problem is probably the most well-studied multiobjective problem. It is the paradigmatic problem for dynamic programming (thus can express a variety of problems), and arises itself directly in many contexts. One area is network routing with various QoS criteria (see e.g. [CX2, ESZ, GR+, VV]). For example, an interesting proposal in a recent paper by Van Mieghen and Vandenberghe [VV] is to have the network operator advertise a portfolio of offered QoS solutions for their network (a trade-off curve), and then users can select the solutions that best fit their applications. Obviously, the portfolio cannot include every single possible route, and it would make sense to select carefully an "optimal" set of solutions that cover well the whole range. Other applications include the transportation of hazardous materials (to minimize risk of accident, and population exposure) [EV], and many others; we refer to the references, e.g. [EG] contains pointers to the extensive literature on shortest paths, spanning trees, knapsack, and the other problems. Our algorithm applies not only to the above standard combinatorial problems, but more generally to any bi-objective problem for which we have available a routine for the Restricted problem; the objective functions and the routine itself could be complex pieces of software without a simple mathematical expression.

After giving the basic definitions and background in Section 2, we present in Section 3 our general lower and upper bound results for bi-objective problems, as well as applications to specific problems. In Section 4 we present some results for $d = 3$ and more objectives. Here we assume only a Gap routine; i.e. these results apply to all problems with a polynomial time constructible $\epsilon$-Pareto set. It was shown in [VY] that for $d = 3$ it is in general impossible to get a constant factor approximation to the optimal $\epsilon$-Pareto set, but one has to relax $\epsilon$. Combining results from [VY] and [KP] we show that for any $\epsilon' > \epsilon$ we can construct an $\epsilon'$-Pareto set of size $c\mathrm{OPT}_\epsilon$, i.e. within a (large) constant factor $c$ of the size $\mathrm{OPT}_\epsilon$ of the optimal $\epsilon$-Pareto set. For general $d$, the problem can be reduced to a Set Cover problem whose VC dimension and codimension are at most $d$, and we can construct an $\epsilon'$-Pareto set of size $O(d \log \mathrm{OPT}_\epsilon)\mathrm{OPT}_\epsilon$.

We discuss also the *Dual* problem: For a specified number $k$ of points, find $k$ points that provide the best approximation to the Pareto curve, i.e. that form an $\epsilon$-Pareto set with the minimum possible $\epsilon$. In [VY] it was shown that for $d = 2$ objectives the problem is NP-hard, but we can approximate arbitrarily well (i.e. there is a PTAS) the minimum approximation ratio $\rho^* = 1 + \epsilon^*$. As we'll see, for $d = 3$ this is not possible, in fact one cannot get any multiplicative approximation (unless P=NP). We use a relationship of the Dual problem to the asymmetric $k$-center problem and techniques from the latter problem to show that the Dual problem can be approximated (for $d = 3$) within a constant power, i.e. we can compute $k$ points that cover every point on the Pareto curve within a factor $\rho' = (\rho^*)^c$ or better in all objectives, for some constant $c$. (It follows from our results that $c \leq 9$.) For small $\rho^*$, i.e. when there is a set of $k$ points that provides a good approximation to the Pareto curve, constant factor and constant power are related, but in general of course they are not.

## 2  Definitions and Background

A multiobjective optimization problem $\Pi$ has a set $\mathcal{I}_\Pi$ of *valid instances*, every instance $I \in \mathcal{I}_\Pi$ has a set of solutions $\mathcal{S}(I)$. There are $d$ objective functions, $f_1, \ldots, f_d$, each of which maps every instance $I$ and solution $s \in \mathcal{S}(I)$ to a value $f_j(I, s)$. The problem specifies for each objective whether it is to be maximized or minimized. We assume as usual in approximation that the objective functions have positive rational values, and that they are polynomial-time computable. We use $m$ to denote the maximum number of bits in numerator and denominator of the objective function values.

We say that a $d$-vector $u$ *dominates* another $d$-vector $v$ if it is at least as good in all the objectives, i.e. $u_j \geq v_j$ if $f_j$ is to be maximized ($u_j \leq v_j$ if $f_j$ is to be minimized). Similarly, we define domination between any solutions according to the $d$-vectors of their objective values. Given an instance $I$, the *Pareto set* $P(I)$ is the set of undominated $d$-vectors of values of the solutions in $\mathcal{S}(I)$. Note that for any instance, the Pareto set is unique. (As usual we are also interested in solutions that realize these values, but we will often blur the distinction and refer to the Pareto set also as a set of solutions that achieve these values. If there is more than one undominated solution with the same objective values, $P(I)$ contains one of them.)

We say that a $d$-vector $u$ *c-covers* another $d$-vector $v$ if $u$ is at least as good as $v$ up to a factor of $c$ in all the objectives, i.e. $u_j \geq v_j/c$ if $f_j$ is to be maximized ($u_j \leq cv_j$ if $f_j$ is to be minimized). Given an instance $I$ and $\epsilon > 0$, an $\epsilon$-*Pareto set* $P_\epsilon(I)$ is a set of $d$-vectors of values of solutions that $(1 + \epsilon)$-cover all vectors in $P(I)$. For a given instance, there may exist many $\epsilon$-Pareto sets, and they may have very different sizes. It is shown in [PY1] that for every multiobjective optimization problem in the aforementioned framework, for every instance $I$ and $\epsilon > 0$, there exists an $\epsilon$-Pareto set of size $O((4m/\epsilon)^{d-1})$, i.e. polynomial for fixed $d$.

An approximate Pareto set always exists, but it may not be constructible in polynomial time. We say that a multiobjective problem $\Pi$ has a polynomial time approximation scheme (respectively a fully polynomial time approximation scheme) if there is an algorithm, which, given instance $I$ and a rational number $\epsilon > 0$, constructs an $\epsilon$-Pareto set $P_\epsilon(I)$ in time polynomial in the size $|I|$ of the instance $I$ (respectively, in time polynomial in $|I|$, the representation size $|\epsilon|$ of $\epsilon$, and in $1/\epsilon$). Let MPTAS (resp. MFPTAS) denote the class of multiobjective problems that have a polynomial time (respectively fully polynomial time) approximation scheme. There is a simple necessary and sufficient condition [PY1], which relates the efficient computability of an $\epsilon$-Pareto set for a multi-objective problem $\Pi$ to the following *GAP Problem*: given an instance $I$ of $\Pi$, a (positive rational) $d$-vector $b$, and a rational $\delta > 0$, either return a solution whose vector dominates $b$ or report that there does not exist any solution whose vector is better than $b$ by at least a $(1 + \delta)$ factor in all of the coordinates. As shown in [PY1], a problem is in MPTAS (resp. MFPTAS) if and only if there is a subroutine GAP that solves the GAP problem for $\Pi$ in time polynomial in $|I|$ and $|b|$ (resp. in $|I|, |b|, |\delta|$ and $1/\delta$).

We say that an algorithm that uses a routine as a black box to access the solutions of the multiobjective problem is *generic*, as it is not geared to a particular problem, but applies to all of the problems for which the particular routine is available. All that such an algorithm needs to know about the input instance is bounds on the minimum and maximum possible values of the objective functions. (For example, if the objective functions are positive rational numbers whose numerators and denominators have at most $m$ bits, then an obvious lower bound on the objective values is $2^{-m}$ and an obvious upper bound is $2^m$; however, for specific problems better bounds may be available.) Based on the bounds, the algorithm calls the given routine for certain values of its parameters, and uses the returned results to compute an approximate Pareto set.

For a given instance, there may exist many $\epsilon$-Pareto sets, and they may have very different sizes. We want to compute one with the smallest possible size, which we'll denote $OPT_\epsilon$. [VY] gives generic algorithms that compute small $\epsilon$-Pareto sets and are applicable to all multiobjective problems in M(F)PTAS, i.e. all problems possessing a (fully) polynomial GAP routine. They consider the following "dual" problems: Given an instance and an $\epsilon > 0$, construct an $\epsilon$-Pareto set of as small size as possible. And dually, given a bound $k$, compute an $\epsilon$-Pareto set with at most $k$ points that has as small an $\epsilon$ value as possible. In the case of two objectives, they give an algorithm that computes an $\epsilon$-Pareto set of size at most $3OPT_\epsilon$; they show that no algorithm can be better than 3-approximate in this setting. For the dual problem, they show that the optimal value of the ratio $\rho = 1 + \epsilon$ can be approximated arbitrarily closely. For three objectives, they show that no algorithm can be $c$-approximate for any constant $c$, unless it is allowed to use a larger $\epsilon$ value. They also give an algorithm that constructs an $\epsilon'$-Pareto set of cardinality at most $4OPT_\epsilon$, for any $\epsilon' > (1 + \epsilon)^2 - 1$.

In a general multiobjective problem we may have both minimization and maximization objectives. In the remainder, we will assume for convenience that all objectives are minimization objectives; this is without loss of generality, since we can simply take the reciprocals of maximization objectives.

*Notation:* For a positive integer $n \in \mathbb{N}^*$, we will denote by $[n]$ the set $\{1, 2, \ldots, n\}$.

## 3 Two Objectives

We use the following notation in this section. Consider the plane whose coordinates correspond to the two objectives. Every solution is mapped to a point on this plane. We use $x$ and $y$ as the two coordinates of the plane. If $p$ is a point, we use $x(p), y(p)$ to denote its coordinates; that is, $p = \big(x(p), y(p)\big)$.

We consider the class of bi-objective problems $\Pi$ for which we can approximately minimize one objective (say the $y$-coordinate) subject to a "hard" constraint on the other (the $x$-coordinate). Our basic primitive is a polynomial time (or fully polynomial time) routine for the following *Restricted problem* (for the $y$-objective): Given an instance $I \in \mathcal{I}_\Pi$, a (positive rational) bound $C$ and a parameter $\delta > 0$, either return a solution point $\tilde{s}$ satisfying $x(\tilde{s}) \le C$ and $y(\tilde{s}) \le (1 + \delta) \cdot \min\{y \text{ over all solutions } s \in \mathcal{S}(I) \text{ having } x(s) \le C\}$ or correctly report that there does not exist any solution $s$ such that $x(s) \le C$. For simplicity, we will drop the instance from the notation and use $\text{Restrict}_\delta\,(y, x \le C)$ to denote the solution returned by the corresponding routine. If the routine does not return a solution, we will say that it returns NO. We say that a routine $\text{Restrict}_\delta\,(y, x \le C)$ runs in polynomial time (resp. fully polynomial time) if its running time is polynomial in $|I|$ and $|C|$ (resp. $|I|, |C|, |\delta|$ and $1/\delta$). The Restricted problem for the $x$-objective is defined analogously. We will also use the Restricted routine with strict inequality bounds; it is easy to see that they are polynomially equivalent.

Note that in general the two objectives could be nonlinear and completely unrelated. Moreover, it is possible that a bi-objective problem possesses a (fully) polynomial Restricted routine for the one objective, but not for the other. The considered class of bi-objective problems is quite broad and contains many well-studied natural ones, most notably the bi-objective shortest path and spanning tree problems (see Section 3.3 for a more detailed list of applications).

The structure of this section is as follows: In Section 3.1, we show that, even if the given bi-objective problem possesses a fully polynomial Restricted routine *for both objectives*, no generic algorithm can guarantee an approximation ratio better than 2. (This lower bound applies *a fortiori* if the Restricted routine is available for one objective only.) Furthermore, we show that for two such natural problems, namely, the bi-objective shortest path and spanning tree problems, it is NP-hard to do better than 2. In Section 3.2 we give a matching upper bound: we present an efficient 2-approximation algorithm that applies to all of the problems that possess a polynomial Restricted routine for one of the two objectives. In Section 3.3 we discuss some applications.

## 3.1 Lower bound

To prove a lower bound for a generic procedure, we present two Pareto sets which are indistinguishable from each other using the Restricted routine as a black box, yet whose smallest $\epsilon$-Pareto sets are of different sizes.

**Proposition 3.1.** *Consider the class of bi-objective problems that possess a fully polynomial Restricted routine for both objectives. Then, for any $\epsilon > 0$, there is no polynomial time generic algorithm that approximates the size of the smallest $\epsilon$-Pareto set $P_\epsilon^*$ to a factor better than 2.*

*Proof.* Fix a rational $\epsilon > 0$ and consider the following set of points: $p = (x(p), y(p))$, $q = \left(x(p)\frac{1+2\epsilon}{1+\epsilon}, \frac{y(p)}{1+\epsilon}\right)$, $r = \left(\frac{x(p)}{1+\epsilon}, y(p)\frac{1+2\epsilon}{1+\epsilon}\right)$, $p_q = \left(x(p)+1, y(p)(1-\frac{1}{x(p)})\right)$ and $p_r = \left(x(p)(1-\frac{1}{y(p)}), y(p)+1\right)$, where $x(p), y(p) > 1 + \frac{1}{\epsilon}$ (Figure 1). Let $P = \{p, q, r, p_q, p_r\}$ and $P' = \{q, r, p_q, p_r\}$ be the feasible (solution) sets corresponding to two input instances. Note that $p$ $(1+\epsilon)$-covers all the points, $p_q$ does not $(1+\epsilon)$-cover $r$ (due to the $x$ coordinate) and $p_r$ does not $(1+\epsilon)$-cover $q$ (due to the $y$ coordinate). It is easy to see that the smallest $\epsilon$ - Pareto set for $P$ consists of only one point (namely point $p$), while the smallest $\epsilon$ - Pareto set for $P'$ *must* include two points.

We can show that a generic algorithm *is guaranteed* to tell the difference between $P$ and $P'$ only if $1/\delta$ is exponential in the size of the input. The argument is very similar to the proof of Theorem 1 in [VY]. Let $x(p) = y(p) = M$, where $M$ is an integer value exponential in the size of the input and $1/\epsilon$. By exploiting the fact that, in some cases, our primitive is *not uniquely defined*, we can argue that a polynomial time generic algorithm cannot distinguish between instances $P$ and $P'$. More specifically, a generic algorithm *is guaranteed* to tell the difference between $P$ and $P'$ only if the tolerance $\delta$ is inverse exponential in the size of the input.

First, note that both points $q$ and $r$ can be efficiently computed by appropriately using the given routine; these two points suffice to $(1 + \epsilon)$-cover the feasible set in both cases. Distinguishing between the two instances means determining whether $p$ is part of the solution. Assume that we use the operation Restrict$_\delta(x, y \leq C)$, where $C \in [y(p), y(p_r)]$. It is easy to see that this is the only "meaningful" operation using this routine as a black box. Then, even if $p$ is part of the solution, by definition, Restrict$_\delta$ can return $p_q$ as long as $x(p_q) \leq (1 + \delta)x(p)$ or equivalently $\delta \geq \frac{1}{M}$. But since we want a polynomial time algorithm, $\frac{1}{\delta}$ has to be polynomial in $\lg M$; hence, the latter constraint must hold. By symmetry, the same property holds for the Restrict$_\delta(y, \cdot)$ routine. Therefore, using each of these routines as a black box, a polynomial time algorithm cannot determine if $p$ is part of the solution, and it is thus forced to take at least two points, even when it is presented with the set $P$. Note that the above configuration can be replicated to show that it is impossible for a generic algorithm to determine whether the smallest $\epsilon$-Pareto set has $k$ points or $2k$ points are needed. ∎

In fact, we can prove something stronger (assuming P $\neq$ NP) for the bi-objective shortest path (*BSP*) and spanning tree (*BST*) problems. In the *BSP* problem, we are given a (directed or undirected) graph, positive rational "costs" and "delays" for each edge and two specified nodes $s$ and $t$. The set of feasible solutions is the set of $s - t$ paths. The objectives (to be minimized) are linear, i.e. the "total weight" of a path equals the sum of the weights of its edges. The *BST* problem is defined analogously. These problems are well-known to possess polynomial Restricted routines for *both* objectives [LR, GR]. We show the following:

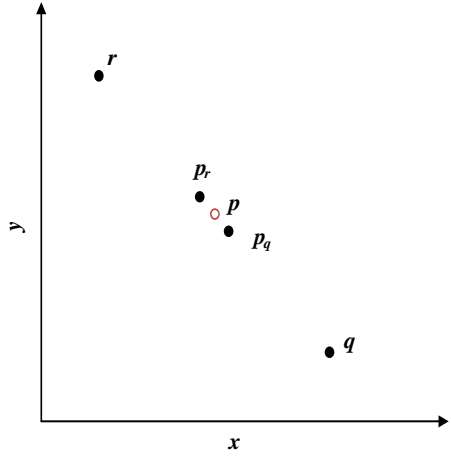

Figure 1: A polynomial time generic algorithm cannot determine if $p$ is a solution of the given instance.

**Theorem 3.2.** *a. For the BSP problem, for any $k$ from $k = 1$ to a polynomial, it is NP-hard to distinguish the case that the minimum size $\mathrm{OPT}_\epsilon$ of the optimal $\epsilon$-Pareto set is $k$ from the case that it is $2k - 1$.*
*b. The same holds for the BST problem for any fixed $k$.*

*Proof.* The reductions are from the Partition problem [GJ]; we are given a set $A$ of $n$ positive integers $A = \{a_1, a_2, \ldots, a_n\}$, and we wish to determine whether it is possible to partition $A$ into two subsets with equal sum.

a. For simplicity, we first prove the theorem for $k = 1$ and then generalize the construction. Given an instance of the Partition problem, we construct an instance of the *BSP* problem as follows: Let $G$ be a graph with $n + 1$ nodes $v_i$, $i \in [n + 1]$ and $2n$ edges $\{e_j, e'_j\}$, $j \in [n]$. We attach the pair of (parallel) edges $\{e_i, e'_i\}$ from $v_i$ to $v_{i+1}$, $i \in [n]$ and set $s \equiv v_1$ and $t \equiv v_{n+1}$. We now specify the two cost functions $c(\cdot)$ and $d(\cdot)$ on the edges: $c(e_i) = d(e'_i) = S + 2\epsilon a_i n$ and $d(e_i) = c(e'_i) = S$, where $S = \sum_{i=1}^n a_i$.

Clearly, this simple transformation defines a bijection between subsets of $[n]$ and $s - t$ paths in $G$; the set $J \subseteq [n]$ is mapped to the $s - t$ path $\mathcal{P}_J = \bigcup_{i \in J}\{e_i\} \cup \bigcup_{i \notin J}\{e'_i\}$. Since $c(\mathcal{P}_J) = nS + 2\epsilon n(\sum_{i \in J} a_i)$ and $d(\mathcal{P}_J) = nS + 2\epsilon n(\sum_{i \notin J} a_i)$, each $s - t$ path $\mathcal{P}$, satisfies the equation $c(\mathcal{P}) + d(\mathcal{P}) = 2(1 + \epsilon)nS$; hence, all feasible solutions are undominated.

Now observe that two solution points suffice to $(1 + \epsilon)$-cover the feasible set; just pick the ("extreme") points $r = ((1 + 2\epsilon)Sn, Sn)$, $l = (Sn, (1 + 2\epsilon)Sn)$, corresponding to the $s - t$ paths $\mathcal{P}_{[n]} = \bigcup_{i=1}^n\{e_i\}$ and $\mathcal{P}_\emptyset = \bigcup_{i=1}^n\{e'_i\}$ respectively. Indeed, $r$ $(1 + \epsilon)$-covers all the points having cost ($x$-coordinate) at least $(1 + \epsilon)Sn$ (since $(1 + 2\epsilon)/(1 + \epsilon) < 1 + \epsilon$). Equivalently, it $(1 + \epsilon)$-covers all the solution points having delay ($y$-coordinate) up to $(1 + \epsilon)Sn$ (since all the solutions lie on the line segment $x + y = 2(1 + \epsilon)nS$). Moreover, the point $l$ $(1 + \epsilon)$-covers all the solution points having $y$-coordinate at least $(1 + \epsilon)Sn$.

Since for each feasible solution $\mathcal{P}$ it holds $\min\{c(\mathcal{P}), d(\mathcal{P})\} \geq nS$ (and the "extreme" paths have cost or delay equal to $nS$), it follows that there exists an $\epsilon$-Pareto set containing (exactly) one point if and only if there exists a path in $G$ with coordinates $((1 + \epsilon)Sn, (1 + \epsilon)Sn)$. It is immediate to verify that such a path exists if and only if there is a solution to the original instance of the Partition problem.

Note that the above part of the proof does not rule out the possibility of an efficient *additive* approximation algorithm, i.e. an algorithm that outputs an $\epsilon$-Pareto set of cardinality at most $\mathrm{OPT}_\epsilon + \alpha$, where $\alpha$ is an absolute constant. We can rule this out as follows: Intuitively, we can think of the Pareto set of $G$ as a "cluster". To prove the theorem for $k > 1$, the goal is to construct an instance of the problem such that the corresponding Pareto set consists of $k$ such clusters that are "$(1 + \epsilon)$-far" from each other, i.e. no point in a cluster $(1 + \epsilon)$-covers any point in a different cluster.

For the *BSP* problem, we can generalize the proof to hold for any $k = \text{poly}(n, \sum_{i=1}^{n} \log(a_i))$ and for all $\epsilon > 0$. This can be achieved by exploiting the combinatorial structure of the problem; we essentially replicate the graph $G$ $k$ times and appropriately scale the weights.

Formally, consider $k$ (disjoint) copies of the graph $G$, $G^j = (V^j, E^j)$, $j \in [k]$, with $V^j = \bigcup_{i=1}^{n+1} \{v_i^j\}$ and $E^j = \bigcup_{i=1}^{n} \{e_i^j, e'_i^j\}$. Add a (source) node $s$, a (sink) node $t$; for each $j$ add an edge from $s$ to $v_1^j$ and one from $v_{n+1}^j$ to $t$. That is, construct the graph $H = (V_H, E_H)$ (see Figure 2) with

$$V_H = \{s, t\} \cup \bigcup_{j=1}^{k} V^j \text{ and } E_H = \bigcup_{j=1}^{k} \{(s, v_1^j) \cup E^j \cup (v_{n+1}^j, t)\}$$

Assign zero cost and delay to each edge incident to $s$ or $t$† and set:

$$
\begin{aligned}
(1 + 2\epsilon)^{2(j-1)} c(e_i^j) &= d(e'_i^j)/(1 + 2\epsilon)^{2(j-1)} = S + 2\epsilon a_i n \\
(1 + 2\epsilon)^{2(j-1)} c(e'_i^j) &= d(e_i^j)/(1 + 2\epsilon)^{2(j-1)} = S
\end{aligned}
$$

From the above equations, it follows that for each $s - t$ path $\mathcal{P}^j$ "using" graph $G^j$, $j \in [k]$, it holds:

$$(1 + 2\epsilon)^{2(j-1)} c(\mathcal{P}^j) + d(\mathcal{P}^j)/(1 + 2\epsilon)^{2(j-1)} = 2(1 + \epsilon)nS$$

This implies that all feasible solutions are undominated. In particular, the Pareto set for this instance is the union of $k$ disjoint "clusters" with endpoints $l_j = \left( \frac{Sn}{(1+2\epsilon)^{2(j-1)}}, Sn(1 + 2\epsilon)^{2(j-1)+1} \right)$ and $r_j = \left( \frac{Sn}{(1+2\epsilon)^{2(j-1)-1}}, Sn(1 + 2\epsilon)^{2(j-1)} \right)$, $j \in [k]$. The solution points in each cluster lie on the line segment $l_j r_j$. (The objective space for this instance is illustrated in Figure 3.)

Now notice that no solution point corresponding to an $s - t$ path using graph $G^j$ is $(1 + \epsilon)$-covered by any point corresponding to an $s - t$ path using graph $G^l$ for $j \neq l$. Indeed, due to the structure of the Pareto set, it suffices to check that, for each $j \in [k - 1]$, the points $l_j$ and $r_{j+1}$ do not $(1 + \epsilon)$-cover each other. This holds by construction: $r_{j+1}$ is a factor of $(1 + 2\epsilon)$ to the left and $(1 + 2\epsilon)$ above $l_j$. Therefore, any two clusters are "$(1 + \epsilon)$-far" from each other. Thus, any $\epsilon$-Pareto set for this instance must contain at least $k$ points.

As in the case of $k = 1$, for all $j \in [k]$, the solution points $l_j$ and $r_j$ $(1 + \epsilon)$-cover the (solution points in the) $j$th cluster. Thus, $2k$ solution points suffice to $(1 + \epsilon)$-cover the feasible set. Also, the $j$th cluster is $(1 + \epsilon)$-covered by one point if and only if there exists an $s - t$ path in $H$ with coordinates $m_j = \left( \frac{(1+\epsilon)Sn}{(1+2\epsilon)^{2(j-1)}}, (1 + \epsilon)Sn(1 + 2\epsilon)^{2(j-1)} \right)$. Similarly, this holds if and only if the original Partition instance is a Yes instance. So, if there exists a partition of the set $A$, the smallest $\epsilon$-Pareto set contains exactly $k$ points. Otherwise, the smallest such set must contain $2k$ points.

To finish the proof, we observe that there exists an $\epsilon$-Pareto set with (at most) $2k - 1$ points if and only if there exists an $\epsilon$-Pareto set with exactly $k$ points. Indeed, the former statement holds if and only if *some* cluster is $(1 + \epsilon)$-covered by *one* point, i.e. if and only if there exists an $s - t$ path in $H$ with coordinates $m_j$ for some $j \in [k]$, which in turn holds if and only if the original Partition instance is a Yes instance. The latter holds if and only if the smallest $\epsilon$-Pareto set contains exactly $k$ points.

b. In the *BST* problem, we are given an undirected graph $G = (V, E)$, positive rational "costs" $c(e)$ and "delays" $d(e)$ for each edge $e \in E$. The set of feasible solutions is the set of spanning trees of $G$; the goal is to minimize cost and delay. For $k = 1$, the proof for the *BST* problem is identical to the proof for the *BSP* problem.

---

†For simplicity, we allow zero weights on the edges, since there does not exist any $s - t$ path with zero total cost or delay. This can be easily removed by appropriate perturbation of the weights.

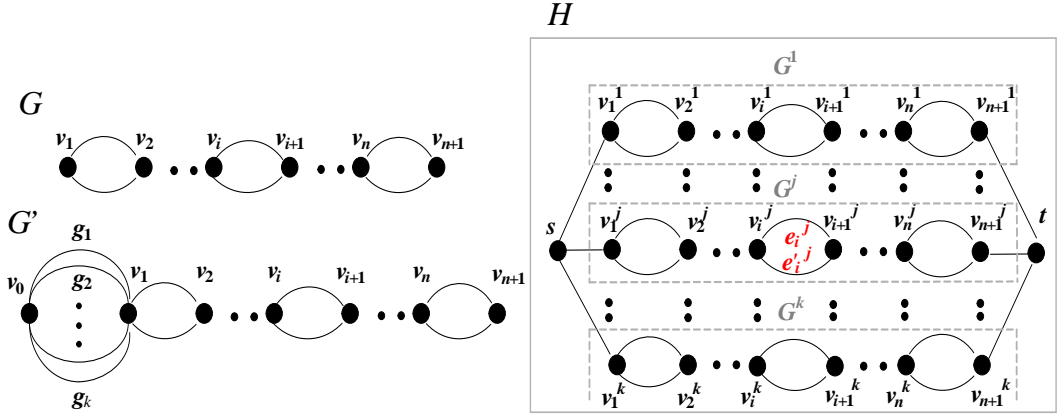

Figure 2: Graphs in the reduction of Theorem 3.2.

We give a construction that works for any *fixed* $k$ and for sufficiently small $\epsilon$; in fact $\epsilon = O(1/k)$ suffices. Consider the graph $G'$ obtained from $G$ by adding one node $v_0$ connected to $v_1$ with $k$ parallel edges $g_i$, $i \in [k]$. Subtract the value $S$ from all the weights of $G$ and set: $c(g_i) = \{2 - (1 + 2\epsilon)^{2i}\} Sn$, $d(g_i) = (1 + 2\epsilon)^{2i} Sn$. (These edges play the role of *offsets*.) Clearly, as long as $(1 + 2\epsilon)^{2k} < 2$, all the weights are in the interval $(0, 2Sn)$. It is also not hard to see that, under this restriction, the Pareto set of $G'$ contains $k$ clusters (having $O(2^n)$ undominated points each) that are "$(1 + \epsilon)$-far" from each other - in the sense indicated above.

The points of the $i$-th cluster, $i \in [k]$, lie on the line segment $c(\mathcal{P}) + d(\mathcal{P}) = 2(1 + \epsilon)nS$ with "endpoints" $(c(g_i), d(g_i) + 2\epsilon Sn)$ and $(c(g_i) + 2\epsilon Sn, d(g_i))$; the latter (solution) points suffice to $(1 + \epsilon)$-cover the corresponding cluster. It is easy to see that there exists a solution with coordinates $((1+\epsilon)c(g_i), (1+\epsilon)d(g_i))$ - i.e. a solution that $(1+\epsilon)$-covers the cluster - if and only if there exists a subset of $A$ with sum $(1+2\epsilon)^{2i}S/2$.

To complete the proof, we use the fact that the following *variant* of the Subset Sum problem is NP-hard: Given $A = \{a_1, a_2, \dots, a_n\}$ with the property that (i) either there exist $k$ subsets $A_i \subseteq A$, $i \in [k]$, such that $\sum_{x \in A_i} x = \gamma^i S/2$ or (ii) *no* such subset exists, decide which one of the two cases holds (for any fixed integer $k$ and rational $\gamma > 1$ such that $\gamma^k < 2$). (This can be shown by a reduction from the Partition problem.) Therefore, it is NP-hard to decide if the smallest $\epsilon$-Pareto set for the instance has $k$ points or $2k$ points are needed. ∎

*Remark* 3.3. For $k = 1$ the theoremark says that it is NP-hard to decide if one point suffices or we need at least 2 points for an $\epsilon$-approximation. We proved that the theorem holds also for more general $k$ to rule out additive and asymptotic approximations. We can easily modify the proof so that the graphs in the reductions are simple. For the *BSP* problem, this can be achieved by inserting a new ("dummy") node in the "middle" of each parallel edge (subdividing the weights arbitrarily). For the *BST* problem, this does not suffice, because all the additional nodes must be covered (by a spanning tree). Let $w_i$ be the node inserted in the middle of $e_i = (v_i, v_{i+1})$. The problem is solved by setting $c((v_i, w_i)) = d((v_i, w_i)) = 0$, $c((w_i, v_{i+1})) = c(e_i)$ and $d((w_i, v_{i+1})) = d(e_i)$. By scaling the weights of the Partition instance we can see that the NP-hardness holds even in the case where all the edge weights are restricted to be positive integers. Similar hardness results can be shown for several other related problems (see Section 3.3).

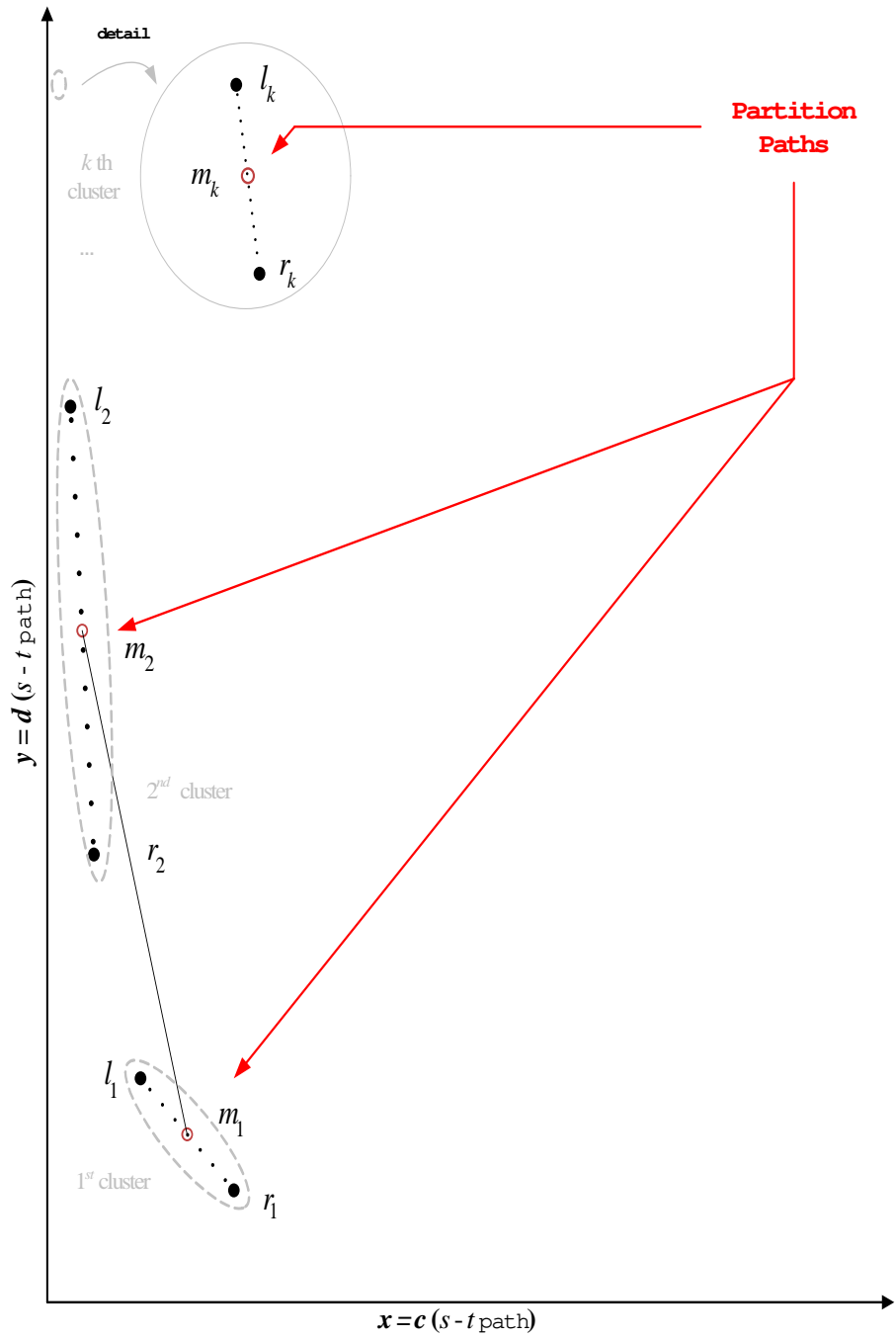

Figure 3: Pareto set for graph $H$ of Theorem 3.2.

## 3.2 Two Objectives Algorithm

We have a bi-objective problem with an associated Restricted routine for the $y$-objective that runs in polynomial (or fully polynomial) time. We are given an instance and an $\epsilon$, and we wish to construct an $\epsilon$-Pareto set of as small size as possible. In this subsection, we present a generic algorithm that guarantees ratio 2. By the result of the previous subsection, this factor is optimal. Recall that the algorithm in [VY] works for all problems in MPTAS and is a factor 3 approximation. (The analysis of the latter algorithm is tight for the class of problems considered here.) In Section 3.2.1, we show that a straightforward greedy approach cannot guarantee a ratio better than 3 in our setting. We next make a crucial observation that is exploited in Section 3.2.2 to achieve the optimal factor.

### 3.2.1 The Greedy Approach Fails

We remark that if the underlying problem has polynomial time *exact* Restricted routines for both objectives (i.e. Restrict$_\delta$ for $\delta = 0$), then we can efficiently compute the *optimal* $\epsilon$-Pareto set by a simple greedy algorithm. The algorithm is similar to the one given in [KP, VY] for the (special) case where all the solution points are given explicitly in the input. We denote by $x_{\min}$, $y_{\min}$ the minimum values of the objectives in each dimension. The greedy algorithm proceeds by iteratively selecting points $q_1, \ldots, q_k$ in decreasing $x$ (increasing $y$) as follows: We start by computing a point $q_1'$ having minimum $y$ coordinate among all feasible solutions (i.e. $y(q_1') = y_{\min}$); $q_1$ is then selected to be the *leftmost* solution point satisfying $y(q_1) \leq (1+\epsilon)y(q_1')$. During the $j$th iteration ($j \geq 2$) we initially compute the point $q_j'$ with minimum $y$-coordinate among all solution points $s$ having $x(s) < x(q_{j-1})/(1+\epsilon)$ and select as $q_j$ the leftmost point which satisfies $y(q_j) \leq (1+\epsilon)y(q_j')$. The algorithm terminates when the last point selected $(1+\epsilon)$-covers the leftmost solution point(s) (i.e. the point(s) $q$ having $x(q) = x_{\min}$). It follows by an easy induction that the set $\{q_1, q_2, \ldots, q_k\}$ is an $\epsilon$-Pareto set of minimum cardinality. (This exact algorithm is applicable to bi-objective linear programming and all problems reducible to it, for example bi-objective flows, the bi-objective *global* min-cut problem [AZ] and several scheduling problems [CJK]. For these problems we can compute an $\epsilon$-Pareto set of minimum cardinality.)

If we have approximate Restricted routines, one may try to modify the greedy algorithm in a straightforward way to take into account the fact that the routines are not exact. However, as shown below, this modified greedy algorithm is suboptimal, in particular it does not improve on the factor 3 that can be obtained from the general GAP routine. More care is required to achieve a factor 2, matching the lower bound.

Suppose that we have a (fully) polynomial Restrict$_\delta$ routine (even for both objectives). Consider the following scheme, where $\delta$ is the "uncertainty parameter" - $\delta < \epsilon$, but $1/\delta$ must be polynomially bounded in the size of the input and $1/\epsilon$, so that the overall algorithm runs in polynomial time:

**Algorithm Greedy**
Compute $y_{\min}$ and $x_{\min}$.
$\bar{y}_1 = y_{\min}(1+\epsilon)$;
$q_1 = \text{Restrict}_\delta(x, y \leq \bar{y}_1)$;
$Q = \{q_1\}$; $i = 1$;
**While** $(x_{\min} < x(q_i)/(1+\epsilon))$ **do**
$\{ q_{i+1}' = \text{Restrict}_\delta(y, x < x(q_i)/(1+\epsilon))$;
$\quad \bar{y}_{i+1} = (1+\epsilon) \cdot \max\{\bar{y}_i, y(q_{i+1}')/(1+\delta)\}$;
$\quad q_{i+1} = \text{Restrict}_\delta(x, y \leq \bar{y}_{i+1})$;
$\quad Q = Q \cup \{q_{i+1}\}$;
$\quad i = i + 1;\ \}$
**Return** $Q$.

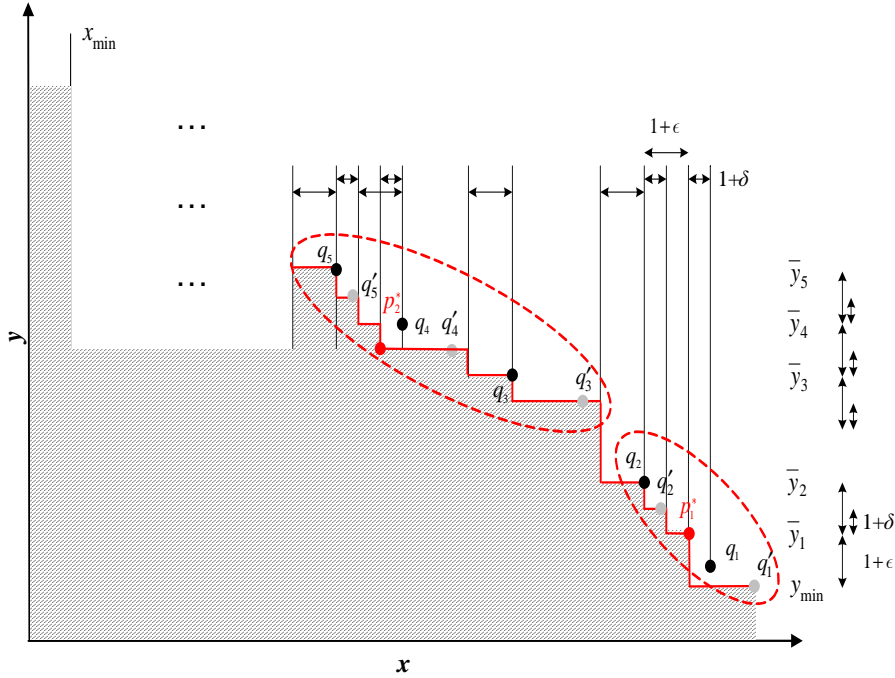

Figure 4: Illustration of the worst-case performance of the greedy approach.

Since the Restricted routines are now approximate, in order to guarantee that the output set of points is an $\epsilon$-Pareto set, we had to appropriately modify the algorithm based on the parameter $\delta$. More specifically, note that the point $q'_{i+1}$ can have $y$-coordinate up to $(1 + \delta)$ times the minimum $y$ over all points $s$ satisfying $x(s) < x(q_i)/(1 + \epsilon)$. In other words, there may exist a solution point $\tilde{s}$ satisfying $x(\tilde{s}) < x(q_i)/(1 + \epsilon)$ and $y(\tilde{s}) = y(q'_{i+1})/(1 + \delta)$. (The algorithm has "no way of knowing this" unless it uses a value of $\delta$ with $1/\delta$ exponential in the size of the input.) This "uncertainty" forces the algorithm to select as point $q_{i+1}$ the leftmost point that satisfies $y(q_{i+1}) \leq (1 + \epsilon)y(q'_{i+1})/(1 + \delta)$. Due to this "weakness", we have the following:

**Claim 3.4.** *For any $\delta > 0$, with $1/\delta$ polynomial in the size of the input and $1/\epsilon$, there exist instances on which the greedy algorithm above outputs a set $Q$ such that $|Q| = 3k - 1$, where $k = \mathrm{OPT}_\epsilon$.*

*Proof.* Denote by $P^*_\epsilon = \{p^*_1, \ldots, p^*_k\}$ the optimal set, where its points $p^*_i$, $i \in [k]$ are ordered in decreasing order of their $x$-coordinate, and $Q = \{q_1, \ldots, q_r\}$ the set selected by the greedy algorithm. By exploiting the *uncertainty* introduced by the parameter $\delta$, we describe an adversarial scenario such that $r = 3k - 1$.

The idea is the following: Consider the subroutine call $q'_{i+1} = \mathrm{Restrict}_\delta(y, x < x(q_i)/(1 + \epsilon))$. By definition, we have $\tilde{y} \leq y(q'_{i+1}) \leq (1 + \delta)\tilde{y}$, where $\tilde{y} = \min\{y(s) \mid x(s) < x(q_i)/(1 + \epsilon)\}$. Suppose that the routine returns a point $q'_{i+1}$ satisfying $\tilde{y} = y(q'_{i+1})$. Call this condition (†). If $q'_{i+1}$ satisfies this condition, the optimal point $p^*_j$ $(1 + \epsilon)$-covering $q'_{i+1}$ can have $y$-coordinate up to $(1 + \epsilon)y(q'_{i+1})$, while the algorithm is forced to select a point $q_{i+1}$ with $y$-value at most $(1 + \epsilon)y(q'_{i+1})/(1 + \delta)$.

We refer the reader to Figure 4 for an illustration. In the instance presented there, the rightmost optimal point $p^*_1$ $(1+\epsilon)$-covers all the solution points that are $(1+\epsilon)$-covered by the set $\{q_1, q_2\}$, while, for $j \geq 2$, the optimal point $p^*_j$ $(1+\epsilon)$-covers all the solution points that are $(1+\epsilon)$-covered by the set $\{q_{3j}, q_{3j+1}, q_{3j+2}\}$. This proves the desired claim. In the following, we explain the situation in detail.

Consider the first point $q_1 \in Q$ selected by the algorithm. By the definition of the Restricted routine and the fact that $q'_1$ must be $(1 + \epsilon)$-covered by $p^*_1$, it follows that $x(p^*_1) \geq x(q_1)/(1 + \delta)$. Suppose that the following scenario occurs: $x(p_1^*) = x(q_1)/(1+\delta)$, $x(q_1)/[(1+\epsilon)(1+\delta)] \leq x(q_2) < x(q_1)/(1+\epsilon)$ and there are no solutions with $x$-coordinate in the interval $[x(q_2)/(1+\epsilon), x(q_2))$. Then, the point $p_1^*$ $(1+\epsilon)$-covers all solutions that are $(1+\epsilon)$-covered by the set $\{q_1, q_2\}$. Notice that the algorithm only "loses" one additional point here; we have that $x(q_2) < x(p_1^*)$. This is due to the fact that we can exactly compute the minimum $y$-coordinate. However, since this does not hold for the next iterations, the algorithm can "lose" two additional points for each optimal point.

Now suppose that the points $\{p_2^*, q_3, q_3', q_4, q_4'\}$ satisfy the following scenario: $q_3'$ satisfies condition (†), $y(q_3) = [(1+\epsilon)/(1+\delta)]y(q_3')$, $y(q_4') = (1+\delta)y(q_3)$, $x(q_4) = (1+\delta)x(p_2^*)$ and $y(p_2^*) = y(q_4')$. It is easy to see that these conditions are simultaneously realizable. (Observe that $p_2^*$ $(1+\epsilon)$-covers $q_3'$.) Finally, if $x(q_4)/[(1+\epsilon)(1+\delta)] \leq x(q_5) < x(q_4)/(1+\epsilon)$ and there are no solutions with $x$-coordinates in the interval $[x(q_5)/(1+\epsilon), x(q_5))$, the point $p_2^*$ $(1+\epsilon)$-covers all the solutions $(1+\epsilon)$-covered by the set $\{q_3, q_4, q_5\}$.

By replicating the above described configuration, it follows inductively that $p_{i+1}^*$ $(1+\epsilon)$-covers all the solutions $(1+\epsilon)$-covered by $\{q_{3i}, q_{3i+1}, q_{3i+2}\}$. This completes the proof. ∎

In fact, one can show that the the greedy algorithm guarantees a factor 3, i.e. the above described adversarial scenario represents a worst-case instance for the algorithm. Let us now try to understand why the greedy approach fails to guarantee a factor 2 in the aforementioned scenario. The problem is that, due to the uncertainty introduced by $\delta$, the point $p_2^*$ can lie arbitrarily to the left of $q_3$. Thus, the only invariant that the greedy algorithm can guarantee is $x(q_4) \leq (1+\delta)x(p_2^*)$.

We can overcome this obstacle by exploiting an additional structural property of the considered class of bi-objective problems. In particular, our generic algorithm will also use a polynomial routine for the following *Dual Restricted* problem (for the $x$-objective): Given an instance, a (rational) bound $D$ and $\delta > 0$, either return a solution $\tilde{s}$ satisfying $y(\tilde{s}) \leq (1+\delta)D$ and $x(\tilde{s}) \leq \min\{x(s)$ over all solutions $s$ having $y(s) \leq D\}$ or correctly report that there does not exist any solution $s$ such that $y(s) \leq D$. Similarly, we drop the instance from the notation and use DualRestrict$_\delta$ $(x, y \leq D)$ to denote the solution returned by the corresponding routine. If the routine does not return a solution, we will say that it returns NO. We say that the corresponding routine runs in polynomial time (resp. fully polynomial time) if its running time is polynomial in $|I|$ and $|D|$ (resp. $|I|$, $|D|$, $|\delta|$ and $1/\delta$).

The following lemma establishes the fact that any bi-objective problem that possesses a (fully) polynomial Restricted routine for the one objective, also possesses a (fully) polynomial *Dual Restricted* routine for the other.

**Lemma 3.5.** *For any bi-objective optimization problem, the problems Restrict$_\delta$ $(y, \cdot)$ and DualRestrict$_\delta$ $(x, \cdot)$ are polynomially equivalent.*

*Proof.* The proof of (both directions of) this equivalence uses binary search on the range of values of one objective with an application of the polynomial routine (for the other objective) at each step of the search. Let $m$ be an upper bound on the number of bits in the objectives; recall that $m$ is polynomially bounded in the size of the instance. Observe that (the absolute value of) the minimum possible difference between the objective values of any two solutions is at least $2^{-2m}$.

First, we argue that a polynomial time algorithm for Restrict$_\delta(y, x \leq C)$ can be used as a black box to obtain a polynomial time algorithm for DualRestrict$_\delta(x, y \leq D)$.

Given an upper bound $D$ and a (rational) error tolerance $\delta > 0$, the following algorithm computes the function DualRestrict$_\delta(x, y \leq D)$:

1. If Restrict$_\delta(y, x \leq 2^m)$ returns a solution $s_0$ having $y(s_0) > (1+\delta)D$ or returns "NO", then output "NO".

2. Otherwise, do a binary search on the parameter $C$ in the range $[2^{-m}, 2^m]$ calling Restrict$_\delta(y, x \leq C)$ in each step, until you find a value $\tilde{C}$ such that:

(a) $\text{Restrict}_\delta(y, x \leq \tilde{C})$ returns a solution $\tilde{s}$ satisfying $x(\tilde{s}) \leq \tilde{C}$ and $y(\tilde{s}) \leq (1+\delta)D$.

(b) $\text{Restrict}_\delta(y, x \leq \tilde{C} - 2^{-2m})$ either returns a solution $s'$ having $x(s') \leq \tilde{C} - 2^{-2m}$ and $y(s') > (1+\delta)D$ or returns "NO".

Output the solution $\tilde{s}$.

The number of calls to the routine $\text{Restrict}_\delta(y, x \leq C)$ is $\Theta(m)$, so the overall algorithm runs in polynomial time. It remains to argue about the correctness. In case 1, either there are no feasible solutions or all solutions have $y$ coordinate strictly greater than $D$. In case 2, all solutions $s$ having $x(s) \leq \tilde{C} - 2^{-2m}$ also satisfy $y(s) > D$. Since there are no solutions with $x$ coordinate strictly between $x(\tilde{s})$ and $\tilde{C} - 2^{-2m}$, $\tilde{C} \leq \min\{x \text{ over all solution points } s \text{ having } y(s) \leq D\}$.

Conversely, given an upper bound $C$ and a (rational) error tolerance $\delta > 0$, the following algorithm computes the function $\text{Restrict}_\delta(y, x \leq C)$ using as a black box an algorithm for $\text{DualRestrict}_\delta(x, y \leq D)$:

1. If $\text{DualRestrict}_\delta(x, y \leq 2^m)$ returns a solution $s_0$ having $x(s_0) > C$ or returns "NO", then output "NO".

2. Otherwise, do a binary search on the parameter $D$ in the range $[2^{-m}, 2^m]$ calling $\text{DualRestrict}_\delta(x, y \leq D)$ in each step, until you find a value $\tilde{D}$ such that:

   (a) $\text{DualRestrict}_\delta(x, y \leq \tilde{D})$ returns a solution $\tilde{s}$ satisfying $x(\tilde{s}) \leq C$ and $y(\tilde{s}) \leq (1+\delta)\tilde{D}$.

   (b) $\text{DualRestrict}_\delta(x, y \leq \tilde{D} - 2^{-2m})$ either returns a solution $s'$ having $x(s') > C$ (and $y(s') \leq (1+\delta)(\tilde{D} - 2^{-2m})$) or returns "NO".

   Output the solution $\tilde{s}$.

The justification is similar. The number of calls to the routine $\text{DualRestrict}_\delta(x, y \leq D)$ is $\Theta(m)$, so the overall running time is polynomial. For the correctness, in case 1, either there are no feasible solutions or all solutions have $x$ coordinate strictly greater than $C$. In case 2, all solutions $s$ having $y(s) \leq \tilde{D} - 2^{-2m}$ also satisfy $x(s) > C$. Since there are no solutions with $y$ coordinate strictly between $y(\tilde{s})$ and $\tilde{D} - 2^{-2m}$, $\tilde{D} \leq \min\{x \text{ over all solution points } s \text{ having } x(s) \leq D\}$. ∎

### 3.2.2 Algorithm Description

We first give a high-level overview of the 2-approximation algorithm. The algorithm iteratively selects a set of solution points $\{q_1, \ldots, q_r\}$ (in decreasing $x$) by judiciously combining the two routines. The idea is, in addition to the Restricted routine (for the $y$-coordinate), to use the Dual Restricted routine (for the $x$-coordinate) in a way that circumvents the problems previously identified for the greedy algorithm. More specifically, after computing the point $q_i'$ in essentially the same way as the greedy algorithm, we proceed as follows. We select as $q_i$ a point that: (i) has $y$-coordinate at most $(1+\epsilon)y(q_i')/(1+\delta)$ and (ii) has $x$-coordinate *at most* the minimum $x$ over all solutions $s$ with $y(s) \leq (1+\epsilon)y(q_i')/(1+\delta)^2$ for a suitable $\delta$. This can be done by a call to the Dual Restricted routine for the $x$-objective. Intuitively this selection means that we give some "slack" in the $y$-coordinate to "gain" some slack in the $x$-coordinate. Also notice that, by selecting the point $q_i$ in this manner, there may exist solution points with $y$-values in the interval $((1+\epsilon)y(q_i')/(1+\delta)^2, (1+\epsilon)y(q_i')/(1+\delta)]$ whose $x$-coordinate is *arbitrarily* smaller than $x(q_i)$. In fact, the optimal point $(1+\epsilon)$-covering $q_i$ can be such a point. However, it turns out that this is sufficient for our purposes and, if $\delta$ is chosen appropriately, this scheme can guarantee that the point $q_{2i}$ lies to the left (or has the same $x$-value) of the $i$-th rightmost point of the optimal solution. We now proceed with the formal description of the algorithm. In what follows, the error tolerance is set to $\delta \doteq \sqrt[3]{1+\epsilon} - 1$ ($\approx \epsilon/3$ for small $\epsilon$). (For the case that the Restricted routine is available for both objectives, we have a variant of this algorithm that achieves a ratio of 2 and is slightly more efficient in the sense that it uses error tolerance $\delta' \doteq \sqrt{1+\epsilon} - 1$.) If $\sqrt[3]{1+\epsilon}$ is not rational, we let $\delta$ be a rational that approximates $\sqrt[3]{1+\epsilon} - 1$ from below, i.e. $(1+\delta)^3 \le (1+\epsilon)$, and which has representation size $|\delta| = O(|\epsilon|)$ (i.e. number of bits in the numerator and denominator). The set of points computed by the algorithm is shown in Figure 5.

**Algorithm 2-Approximation**
**If** $\text{Restrict}_{\delta_0 \leftarrow 1}(y, x \le 2^m) = \text{NO}$ **then** halt.
$q'_1 = \text{Restrict}_\delta(y, x \le 2^m)$;
$q_{\text{left}} = \text{DualRestrict}_{\delta_0 \leftarrow 1}(x, y \le 2^m)$;    $x_{\min} = x(q_{\text{left}})$;
$\bar{y}_1 = y(q'_1)(1+\delta)$;
$q_1 = \text{DualRestrict}_\delta(x, y \le \bar{y}_1)$;
$\bar{x}_1 = x(q_1)/(1+\epsilon)$;
$Q = \{q_1\}; i = 1$;
**While** ($\bar{x}_i > x_{\min}$) **do**
$\{\ q'_{i+1} = \text{Restrict}_\delta(y, x < \bar{x}_i)$;
    $\bar{y}_{i+1} = [(1+\epsilon)/(1+\delta)] \cdot \max\{\bar{y}_i, y(q'_{i+1})/(1+\delta)\}$;
    $q_{i+1} = \text{DualRestrict}_\delta(x, y \le \bar{y}_{i+1})$;
    $\bar{x}_{i+1} = x(q_{i+1})/(1+\epsilon)$;
    $Q = Q \cup \{q_{i+1}\}$;
    $i = i + 1; \}$
**Return** $Q$.

### 3.2.3 Analysis

Recall that $2^m$ is an upper bound on the values of the objectives. Thus, if $\text{Restrict}_{\delta_0 \leftarrow 1}(y, x \le 2^m) = \text{NO}$, there are no feasible solutions, in which case we can just terminate the algorithm. So, we can assume that the solution set is nonempty. In this case, the subroutine calls of lines 2 and 3 indeed return a solution; moreover, (*i*) the solution point $q_{\text{left}}$ has minimum $x$-value among all feasible solutions and (*ii*) $q'_1$ has $y$-value *at most* $(1+\delta)y_{\min}$. Now observe that $y_{\min} \le \bar{y}_i \le \bar{y}_{i+1}$ and $\bar{x}_i > x_{\min}$ for all the values of $i$ for which the body of the while loop is executed. It is thus easy to see that each subroutine call returns a point; so, all the points are well-defined.

Let $Q = \{q_1, q_2, \dots, q_r\}$ be the set of solution points produced by the algorithm. We will prove that the set $Q$ is an $\epsilon$-Pareto set whose size is at most twice the optimum. We note the following simple properties.

**Fact 3.6.** *We have the following:*
*1. For each $i \in [r-1]$ it holds (i) $x(q'_{i+1}) < x(q_i)/(1+\epsilon)$ and (ii) for each solution point $t$ with $x(t) < x(q_i)/(1+\epsilon)$, we have $y(t) \ge y(q'_{i+1})/(1+\delta)$.*
*2. For each $i \in [r]$ it holds (i) $y(q_i) \le (1+\delta)\bar{y}_i$ and (ii) for each solution point $t$ with $y(t) \le \bar{y}_i$ we have $x(t) \ge x(q_i)$.*

*Proof.* The properties are just restatements of the definition of the two subroutines. ∎

We can now prove the following lemmata (all properties used below refer to the above fact).

**Lemma 3.7.** *The $x$ coordinates of the points $q_1, q_2, \dots, q_r$ of $Q$ form a strictly decreasing sequence.*

*Proof.* Consider two successive elements $q_i, q_{i+1}$ of $Q$. For their $x$ coordinates we will argue that $x(q_{i+1}) < x(q_i)/(1+\epsilon)$. First observe that $y(q'_{i+1}) \le \bar{y}_{i+1}$. So, property 2-(*ii*) implies that $x(q_{i+1}) \le x(q'_{i+1})$. Now from property 1-(*i*) we get $x(q'_{i+1}) < x(q_i)/(1+\epsilon)$ and the argument is complete. ∎

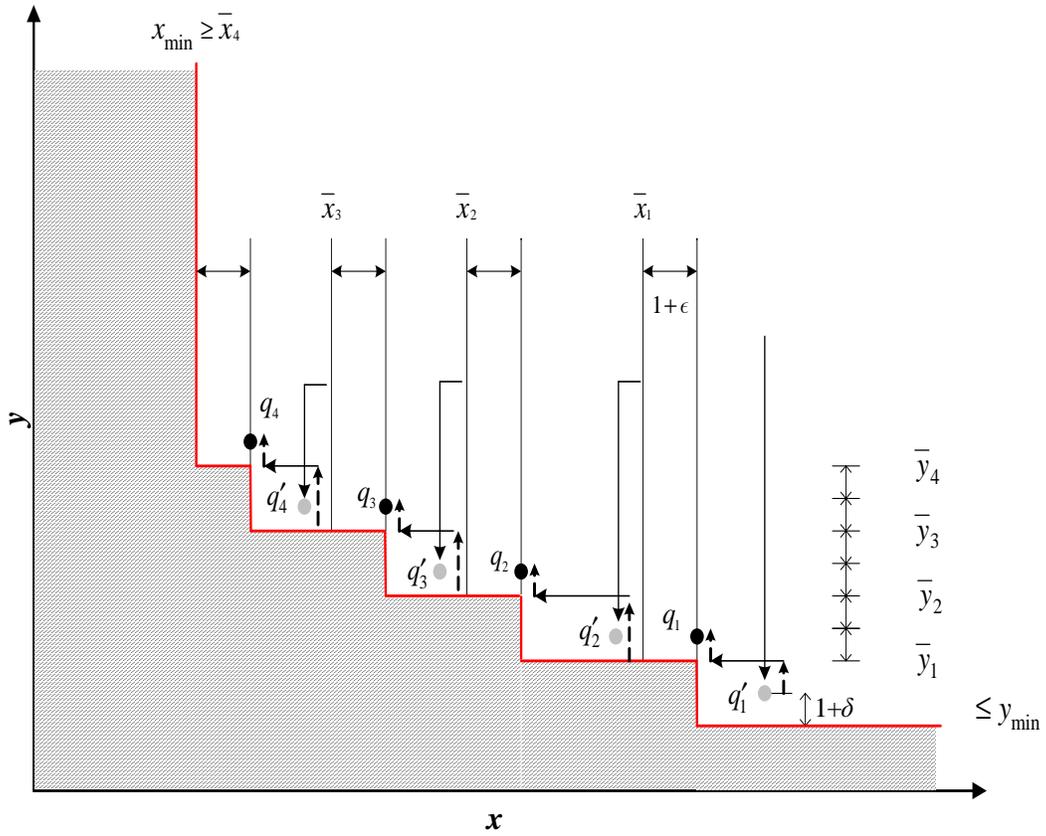

Figure 5: Schematic performance of the algorithm. The scale is logarithmic in both dimensions. There are no solutions in the shaded region.

The following lemma shows that $Q$ is indeed an $\epsilon$-Pareto set.

**Lemma 3.8.** *1. The point $q_1$ $(1 + \epsilon)$-covers all of the solution points that have $x$-coordinate at least $x(q_1)/(1+\epsilon)$.*
*2. For each $i \in [r] \setminus \{1\}$ the point $q_i$ $(1+\epsilon)$-covers all of the solution points that have their $x$-coordinate in the interval $\big[ x(q_i)/(1+\epsilon), x(q_{i-1})/(1+\epsilon) \big)$.*
*3. There are no solution points with $x$-coordinate smaller than $x(q_r)/(1+\epsilon)$.*

*Proof.* 1. Let $t$ be a solution point with $x(t) \geq x(q_1)/(1+\epsilon)$. We need to show that $t$ is $(1+\epsilon)$-covered by $q_1$. It clearly suffices to argue that $y(t) \geq y(q_1)/(1+\epsilon)$. Indeed, by property 2-*(ii)* we have $y(q_1) \leq (1+\delta)\bar{y}_1 = (1+\delta)^2 y(q'_1)$ and the definition of $q'_1$ implies that $y(t) \geq y(q'_1)/(1+\delta)$, *for any solution point $t$.* By combining these facts we get that for any solution point $t$ it holds $y(t) \geq y(q_1)/(1+\delta)^3 \geq y(q_1)/(1+\epsilon)$.
2. Let $t$ be a solution point satisfying $x(q_i)/(1+\epsilon) \leq x(t) < x(q_{i-1})/(1+\epsilon)$; we will show that $t$ is $(1+\epsilon)$-covered by $q_i$ or equivalently that $y(t) \geq y(q_i)/(1+\epsilon)$. The proof is by contradiction. Suppose that there exists such a point $t$ with $y(t) < y(q_i)/(1+\epsilon)$. By property 2-*(i)* and the definition of $\bar{y}_i$ this implies $y(t) < \max\{\bar{y}_{i-1}, y(q'_i)/(1+\delta)\}$. Now since $x(t) < x(q_{i-1})/(1+\epsilon)$, property 1-*(ii)* gives $y(t) \geq y(q'_i)/(1+\delta)$. Furthermore, since $x(t) < x(q_{i-1})$, by property 2-*(ii)* it follows that $y(t) > \bar{y}_{i-1}$. This provides the desired contradiction.
3. The termination condition of the algorithm is $x(q_r)/(1+\epsilon) \leq x_{\min}$. ∎

*Remark* 3.9. We show in Lemma 3.10 below that the set $Q$ is of cardinality $|Q| \leq 2\mathrm{OPT}_\epsilon$. So, the algorithm could output this set of points. However, we observe that the set $Q$ may contain "redundant" points: The $y$-coordinates of the points $q_1, \ldots, q_r$ do not necessarily form an increasing sequence. In fact, if $y(q_{i+1}) \leq (1+\delta)\bar{y}_i$, it may happen that $y(q_{i+1}) \leq y(q_i)$ (in which case the point $y_i$ is redundant). (Note however that if $y(q_{i+1}) > (1+\delta)\bar{y}_i$, then by property 2-*(i)* we get $y(q_{i+1}) > y(q_i)$.) This observation can be further exploited for a post-processing step. For example, if $y(q_{2i}) \leq (1+\delta)\bar{y}_{2i-1}$, we can safely discard the point $q_{2i-1}$ as implied by (the proof of) Lemma 3.8.

We now bound the size of the set of points $Q$ in terms of the size of the optimal $\epsilon$-Pareto set.

**Lemma 3.10.** *Let $P^*_\epsilon = \{p^*_1, p^*_2, \ldots, p^*_k\}$ be the optimal $\epsilon$-Pareto set, where its points $p^*_i$, $i \in [k]$, are ordered in (strictly) increasing order of their $y$- and (strictly) decreasing order of their $x$-coordinate. Then, $|Q| = r \leq 2k$.*

*Proof.* We prove the following:

**Claim 3.11.** *If the algorithm selects a solution point $q_{2i-1}$ (i.e. if $2i - 1 \leq r$), then there must exist a point $p^*_i$ in $P^*_\epsilon$ (i.e. it holds $i \leq k$) and if the algorithm selects a point $q_{2i}$, then $x(p^*_i) \geq x(q_{2i})$.*

The desired result follows directly from this. The claim is proved by induction on $i$.

Basis $(i = 1)$. The first statement of the claim trivially holds. To show the validity of the second statement observe that for the rightmost point of $P^*_\epsilon$, we must have $y(p^*_1) \leq y(q'_1)(1+\epsilon) = \bar{y}_1(1+\epsilon)/(1+\delta) \leq \bar{y}_2$. The first inequality holds since the solution point $q'_1$ must be $(1+\epsilon)$-covered by $P^*_\epsilon$ and in particular by the point of $P^*_\epsilon$ having the minimum $y$-coordinate. The two other inequalities follow from the definitions of $\bar{y}_1$ and $\bar{y}_2$. Now an application of property 2-*(ii)* gives $x(p^*_1) \geq x(q_2)$ and the base case is proved.

Induction step. Suppose that the claim holds for index $i-1$ (more specifically that $x(p^*_{i-1}) \geq x(q_{2i-2})$); we will prove it for $i$. We will prove each statement in turn.

Assume first that the algorithm selects a point $q_{2i-1}$ (i.e. that $2i - 1 \leq r$). We will show that $P^*_\epsilon$ contains a point $p^*_i$ (i.e. that $i \leq k$). By the termination condition of the algorithm, our assumption implies that $x(q_{2i-2}) > (1+\epsilon)x_{\min}$. Therefore, by the induction hypothesis it follows that $x(p^*_{i-1}) > (1+\epsilon)x_{\min}$; that is, point $p^*_{i-1}$ does *not* $(1+\epsilon)$-cover the leftmost solution point, which means there must exist a point $p^*_i$ in the optimal set.

Now assume that the algorithm selects a point $q_{2i}$. We will show that $x(p^*_i) \geq x(q_{2i})$. First note that by property 1-($i$) and the induction hypothesis $x(q'_{2i-1}) < x(p^*_{i-1})/(1+\epsilon)$. So, the point $p^*_{i-1}$ does *not* $(1+\epsilon)$-cover the point $q'_{2i-1}$ in the $x$-coordinate. Clearly, the latter point must be $(1+\epsilon)$-covered by a point in $P^*_\epsilon$. Since the $p^*_j$'s are sorted in decreasing order of their $x$-coordinates, we conclude that $p^*_i$ is the only eligible point for that purpose, i.e. $q'_{2i-1}$ must be must $(1+\epsilon)$-covered by $p^*_i$. To complete the argument, we need the following fact:

**Fact 3.12.** *There does not exist any solution point $t$ with $x(t) < x(q_{2i})$ such that $t$ $(1+\epsilon)$-covers point $q'_{2i-1}$.*

*Proof.* We want to prove that for all solutions $t$ having $x(t) < x(q_{2i})$ it holds $y(t) > (1+\epsilon)y(q'_{2i-1})$. For such a solution point $t$ we have $y(t) > \bar{y}_{2i} \geq \bar{y}_{2i-1}(1+\epsilon)/(1+\delta) \geq (1+\epsilon)y(q'_{2i-1})$. The latter inequalities, in the order they appear, follow by applying property 2-($ii$) and the definition of $\bar{y}_j$ (for $j = 2i-1, 2i$). ∎

The above fact implies directly that $x(p^*_i) \geq x(q_{2i})$ and the proof is complete. ∎

Thus far, we have proved that the set $Q$ is an $\epsilon$-Pareto set of size $|Q| \leq 2\text{OPT}_\epsilon$. We now analyze the running time of the algorithm. Let $k$ be the number of points in the smallest $\epsilon$-Pareto set, $k = \text{OPT}_\epsilon$. The algorithm involves $r \leq 2k$ iterations of the while loop; each iteration involves two calls to the subroutines. Therefore, the total running time is bounded by $4k$ subroutine calls. In summary, we proved the following theorem.

**Theorem 3.13.** *The above described algorithm computes a 2-approximation to the smallest $\epsilon$-Pareto set in time $O(\text{OPT}_\epsilon)$ subroutine calls, where $1/\delta = O(1/\epsilon)$.*

## 3.3 Applications

Our result can be applied to all of the problems which have a polynomial (or fully polynomial) time Restricted routine for one of the two objectives. It should be stressed that our algorithm is quite general; it does not assume for example linearity of the objectives. Applications include the shortest path problem [Han, Wa, ESZ, LR] and generalizations [EV, GR+, CX2, VV], cost-time trade-offs in query evaluation [PY2], spanning trees (and more generally matroid problems, see below) [GR, HL] and related problems [CX]. The aforementioned problems possess a polynomial Restricted routine for *both* objectives. In essence, for most of the aforementioned problems (with [PY2] being a notable exception), the two objectives are "the same" and we can efficiently optimize each of them separately. For several other problems [ABK1, ABK2, CJK, DJSS], the Restricted routine is available for one objective *only* (because it is NP-hard to separately optimize this objective). An example is the following classical scheduling problem: We are given a set of $n$ jobs and a fixed number $m$ of machines. Executing job $j$ on machine $i$ requires time $p_{ij}$ and incurs cost $c_{ij}$. We are interested in the trade-off between makespan and cost. Minimizing the makespan is NP-hard, even for $m = 2$; hence, the Dual Restricted problem for this objective (equivalently, the Restricted problem for the cost objective) does not have a PTAS. If $m$ is fixed, a fully polynomial time *Dual Restricted* routine for the cost objective is given in [ABK1]. (By Lemma 3.5 this implies an FPTAS for the Restricted problem for the makespan objective.)

For the bi-objective shortest path problem, a polynomial (resp. fully polynomial) Restricted routine corresponds to a polynomial (resp. fully polynomial) time approximation scheme for the *Restricted Shortest Path* problem: given a bound on the cost of the path, minimize the delay of the path subject to the bound on the cost. This problem has been studied in a number of papers [Has, Wa, LR, ESZ]. The problem is NP-hard and has a fully polynomial time approximation scheme. The best current algorithms approximate the optimal restricted path within factor $1 + \epsilon$ in time $O(en/\epsilon)$ for acyclic (directed) graphs [ESZ], and time $O(en(\log \log n + 1/\epsilon))$ for general (directed) graphs [LR], where $n$ is the number of nodes and $e$ is the number of edges. Moreover, the Dual Restricted problem also admits an FPTAS with the same time complexity. Thus, our algorithm runs in $O(en(\log \log n + 1/\epsilon)\mathrm{OPT}_\epsilon)$ time for general graphs and $O(en\mathrm{OPT}_\epsilon/\epsilon)$ for acyclic graphs. The time complexity is comparable or better than previous algorithms [Han, Wa, TZ], which furthermore do not provide any guarantees on the size.

For the bi-objective spanning tree problem a polynomial Restricted routine corresponds to a polynomial time approximation scheme for the *Constrained Spanning Tree* (*CST*) problem [GR]: given a bound on the cost of the tree, minimize the delay of the tree subject to the bound on the cost. This problem is also NP-hard and is known to have a polynomial time approximation scheme [GR, HL]. (In fact, the aforementioned papers provide a PTAS for the more general problem of finding a minimum cost base of a matroid subject to a bound on the total length, as long as there is a polynomial time independence oracle for the matroid.) The best current algorithm for the problem [HL] has running time $O((1/\epsilon)^{1/\epsilon}n^3)$. As a corollary, our generic algorithm can compute a 2-approximation to the smallest $\epsilon$-Pareto set in time $O((1/\epsilon)^{1/\epsilon}n^3\mathrm{OPT}_\epsilon)$. Whether such a 2-approximation can be computed in *fully* polynomial time is conditional on the existence of an FPTAS for the *CST* problem (which is an interesting open question). In contrast, by the results of [PY1, VY], a 3-approximation can be computed in fully polynomial time.

# 4  $d$ Objectives

The results in this section use the GAP routine and thus apply to all problems in MPTAS.

## 4.1  Approximation of the optimal $\epsilon$-Pareto set.

Recall that for $d \geq 3$ objectives we are forced to compute an $\epsilon'$-Pareto set, where $\epsilon' > \epsilon$, if we are to have a guarantee on its size [VY]. For any $\epsilon' > \epsilon$, a logarithmic approximation for the problem is given in [VY], by a straightforward reduction to the Set Cover problem. We can sharpen this result, by exploiting additional properties of the corresponding set system.

**Theorem 4.1.** *1. For any $\epsilon' > \epsilon$ there exists a polynomial time generic algorithm that computes an $\epsilon'$-Pareto set $Q$ such that $|Q| \leq O\big(d \log \mathrm{OPT}_\epsilon\big)\mathrm{OPT}_\epsilon$. The algorithm uses $O((m/\delta)^d)$ $\mathrm{GAP}_\delta$ calls, where $1/\delta = O(1/(\epsilon' - \epsilon))$.*
*2. For $d = 3$, the algorithm outputs an $\epsilon'$-Pareto set $Q$ satisfying $|Q| \leq c\mathrm{OPT}_\epsilon$, where $c$ is a constant.*

Consider the following problem $\mathcal{Q}(P, \epsilon)$: Given a set of $n$ points $P \subseteq \mathbb{R}_+^d$ as input and $\epsilon > 0$, compute the smallest $\epsilon$-Pareto set of $P$. It should be stressed that, by definition, the set of points $P$ is given *explicitly* in the input. (Note the major difference with our setting: for a typical multiobjective problem there are exponentially many solution points and they are not given explicitly.) This problem can be solved in linear time for $d = 2$ by a simple greedy algorithm. For $d = 3$ it is NP-hard and can be approximated within some (large) constant factor $c$ [KP]. If $d$ is arbitrary (i.e. part of the input, e.g. $d = n$), the problem is hard to approximate better than within a $\Omega(\log n)$ factor (unless P = NP) [VY].

The following simple fact, implicit in [VY], relates the approximability of $\mathcal{Q}$ with the problem of computing a small $\epsilon'$-Pareto set for a multiobjective problem $\Pi$, given the GAP primitive. Let $\epsilon > 0$ be a given rational number. For any $\epsilon' > \epsilon$, we can find a $\delta > 0$ such that $1/\delta = O(1/(\epsilon' - \epsilon))$ satisfying $1 + \epsilon' \geq (1 + \epsilon)(1 + \delta)^2$.

**Lemma 4.2.** *Suppose that there exists an $r$-factor approximation algorithm for $\mathcal{Q}$. Then, for any $\epsilon' > \epsilon$, we can compute an $\epsilon'$-Pareto set $Q$, such that $|Q| \leq r\mathrm{OPT}_\epsilon$ using $O((m/\delta)^d)$ $\mathrm{GAP}_\delta$ calls.*

*Proof.* The algorithm proceeds in two phases; in the first phase, we compute a $\delta$-Pareto set, by using the original algorithm of [PY1] and in the second phase we post-process the points produced by the latter algorithm by using the $r$-approximation algorithm for $\mathcal{Q}$ as a black box.

For the given instance $I \in \mathcal{I}_\Pi$, let $\mathfrak{X}(I)$ be the set of $d$-vectors of values of solutions in the objective space and fix an optimal $\epsilon$-Pareto set $P_\epsilon^* = P_\epsilon^*(I)$. Let $R$ be the $\delta$-Pareto set produced in the first stage. We apply the $r$-approximation algorithm for $\mathcal{Q}$ on input $R$ to produce a set $R' \subseteq R$ that $(1+\epsilon)(1+\delta)$-covers $R$. (Since $|R| \leq (m/\delta)^{d-1}$, it follows that the overall algorithm runs in polynomial time.) $R'$ is clearly an $\epsilon'$-Pareto set for the feasible set $\mathfrak{X}(I)$. We will argue that $|R'| \leq r\mathrm{OPT}_\epsilon$. Let $R^*$ denote the smallest $(1+\epsilon)(1+\delta)$-cover for $R$ using *only* points *from* $R$; we have $|R'| \leq r|R^*|$. The following simple claim completes the argument:

**Claim 4.3.** $|R^*| \leq \mathrm{OPT}_\epsilon$.

*Proof.* It suffices to show that there exists an $(1+\epsilon)(1+\delta)$-cover $C$ for $R$ of cardinality at most $\mathrm{OPT}_\epsilon$. Since $R$ is a $\delta$-Pareto set, for any solution point $s \in \mathfrak{X}(I)$, there exists a solution point $r \in R$ that $(1+\delta)$-covers $s$. $C$ is constructed as follows: For each $s \in P_\epsilon^*$ pick an $r \in R$ that $(1+\delta)$-covers it. Then, $|C| \leq |P_\epsilon^*| = \mathrm{OPT}_\epsilon$. Every point $r \in R$ is $(1+\epsilon)$-covered by a point $s \in P_\epsilon^*$, which in turn is $(1+\delta)$-covered by a point $c \in C$. Therefore, $C$ $(1+\epsilon)(1+\delta)$-covers all points of $R$. ■ ■

Part 2 of Theorem 4.1 follows immediately from the fact that $\mathcal{Q}$ is constant factor approximable for $d = 3$ [KP] and Lemma 4.2. We consider the case of general $d$ in the remainder. To proceed, we need the following definition.

**Definition 4.4.** A set system is a pair $(U, \mathcal{R})$, where $U$ is a set and $\mathcal{R}$ is a collection of subsets of $U$. For a set system $(U, \mathcal{R})$, we say that $X \subseteq U$ is *shattered* by $\mathcal{R}$ if for any $Y \subseteq X$, there exists a set $R \in \mathcal{R}$ with $X \cap R = Y$. The VC-dimension [VC] of the set system is the maximum size of any set shattered by $\mathcal{R}$. Let $T \subseteq U$ be a finite set and $r \in (1, \infty)$ be a parameter. A set $N \subseteq T$ is called an $1/r$-net for $(T, \mathcal{R})$ [HW], if $N \cap S \neq \emptyset$ for all $S \in \mathcal{R}$ having $|S| > |T|/r$.

The problem $\mathcal{Q}(P, \epsilon)$ can be formulated as a set cover problem as follows: For each point $q \in P$ and $\epsilon > 0$, define $S_{q,\epsilon} = \{x \in \mathbb{R}^d \mid q \leq (1+\epsilon) \cdot x\}$. $S_{q,\epsilon}$ is the subset of $\mathbb{R}^d$ that is $(1+\epsilon)$-covered by $q$; it is a closed convex cone in $\mathbb{R}^d$ (a translation of the nonnegative orthant by the vector $q/(1+\epsilon)$). For each point $r \in P$, $r$ is $(1+\epsilon)$-covered by $q$ if and only if $r \in S_{q,\epsilon}$. Now consider the set system $\mathcal{F}(P, \epsilon) = (P, \mathcal{S}(P, \epsilon))$, where $\mathcal{S}(P, \epsilon) = \{P_{q,\epsilon} \equiv P \cap S_{q,\epsilon} \mid q \in P\}$. Clearly, there is a bijection between set covers of $\mathcal{F}(P, \epsilon)$ and $\epsilon$-Pareto sets of $P$. We now establish the following:

**Lemma 4.5.** *a. For any finite set of points $P \subseteq \mathbb{R}^d$ and $\epsilon > 0$, it holds VC-dim$(\mathcal{F}(P, \epsilon)) \leq d$.*
*b. There exists a set of points $P$ such that VC-dim$(\mathcal{F}(P, \epsilon)) = d$.*

*Proof.* a. Let $P$ be a set of points in $\mathbb{R}^d$ and $\epsilon > 0$. We must argue that *no* subset $P' \subseteq P$ of cardinality $d + 1$ can be shattered by $\mathcal{S}(P, \epsilon)$. Note that any such set $P' \subseteq P$ (of cardinality $d + 1$) contains a point $r$ none of whose coordinates is minimal, that is, a point $r$ such that for all $i \in [d]$ there exists some point $q^i \in P'$ (different from $r$) with the property $(q^i)_i \leq r_i$. We claim that we cannot "separate" $r$ from the remaining points of $P'$ by any convex cone (as defined above). Indeed, a point that $(1+\epsilon)$-covers the $q^i$'s is guaranteed to $(1+\epsilon)$-cover $r$ (or equivalently, the "dichotomy" $\{q^i, i \in [d]\}$ cannot be realized).

b. Consider a set $P = A \cup C$, where $|A| = d$ and $|C| = 2^d$. Let $A = \{a_1, \ldots, a_d\}$. We select the $a_i$' s in $A$ as follows: For each $i \in [d]$, the $i$th coordinate of $a_i$ is equal to 1 and all the rest are equal to $1 + 2\epsilon$. The set $A$ has two properties: (i) no two of its points $(1 + \epsilon)$-cover each other and (ii) for any two points $p, q \in A$, we have $\mathrm{argmin}_i p_i \neq \mathrm{argmin}_i q_i$. The set $C$ is selected such that each subset of $A$ is $(1 + \epsilon)$-covered by some point in $C$. In particular, let $X = \bigcup_{i \in \mathcal{I}(X)} \{a_i\}$ be a subset of $A$. We add the point $c_X$ in $C$ having each coordinate indexed by $\mathcal{I}(X)$ equal to $1 + \epsilon$ and all the rest equal to $1 + 2\epsilon$. Clearly, the point $c_X$ $(1 + \epsilon)$-covers *exactly* the elements of $X$. ∎

For $q \in P$ and $\epsilon > 0$, define $S_{q,\epsilon}^D = \{x \in \mathbb{R}^d \mid x \leq (1 + \epsilon) \cdot q\}$; the cone $S_{q,\epsilon}^D$ is the subset of $\mathbb{R}^d$ that $(1 + \epsilon)$-covers $q$. A point $r$ $(1 + \epsilon)$-covers $q$ if and only if $r \in S_{q,\epsilon}^D$. The "dual" set system of $\mathcal{F}(P, \epsilon)$ is defined as $\mathcal{F}^D(P, \epsilon) = (P, \mathcal{S}^D(P, \epsilon))$, where $\mathcal{S}^D(P, \epsilon) = \{P_{q,\epsilon}^D \equiv P \cap S_{q,\epsilon}^D \mid q \in P\}$. In words, the elements are the points of $P$ and for each point $q \in P$ we have a set consisting of the points $r \in P$ that $(1 + \epsilon)$-cover $q$. An $\epsilon$-Pareto set of $P$ is equivalent to a hitting set of $\mathcal{F}^D$.

It is well-known [As] that, if a set system has VC-dimension at most $d$, the VC-dimension of the dual set system is upper bounded by $2^{d+1} - 1$. However, in our setting, essentially the same proof as in the previous lemma establishes the following:

**Lemma 4.6.** *For any finite set of points $P \subseteq \mathbb{R}^d$ and $\epsilon > 0$, it holds VC-dim($\mathcal{F}^D(P, \epsilon)$) $\leq d$. This bound is tight.*

*Proof.* Let $P$ be a set of points in $\mathbb{R}^d$ and $\epsilon > 0$. We must argue that *no* subset $P' \subseteq P$ of cardinality $d + 1$ can be shattered by $\mathcal{S}^D(P, \epsilon)$. Similarly to the previous lemma, any set $P' \subset P$ of cardinality $d+1$ contains a point $r$ such that for all $i \in [d]$ there exists some point $q^i \in P'$ ($q^i \neq r$) satisfying $(q^i)_i \geq r_i$. We claim that we cannot "separate" $r$ from the remaining points. Indeed, if some point is $(1 + \epsilon)$-covered by all the $q^i$'s, then is also $(1 + \epsilon)$-covered by $r$. The tightness is similar. ∎

It is well-known that, for a set system of VC-dimension at most $d$, we can efficiently construct an $1/r$-net of size $s(r) = O(dr \log r)$ [KPW]; this bound is tight in general [PW, KPW]. As shown in [BG, ERS], for such a set system, there exists a polynomial time $s(\mathrm{OPT})/\mathrm{OPT}$-factor approximation algorithm for the minimum *hitting set* problem, where OPT is the cost of the optimal solution. If we apply this result to the dual set system $\mathcal{F}^D(P, \epsilon)$ we conclude:

**Proposition 4.7.** *Problem $\mathcal{Q}$ can be approximated within a factor of $O(d \log \mathrm{OPT}_\epsilon)$.*

Part 1 of Theorem 4.1 follows by combining Lemma 4.2 and Proposition 4.7.

*Remark* 4.8. If $s(r) = O(r)$, the reduction in [BG, ERS] implies a polynomial time constant factor approximation algorithm for the corresponding hitting set problem. This is exactly the approach in [KP]: they show that, for $d = 3$, $\mathcal{F}^D(P, \epsilon)$ admits an $1/r$-net of size $s(r) = O(r)$ and that such a net can be efficiently constructed. Note that the constant approximation ratio $c$ implied for set cover using this approach is identified with the constant hidden in the big-Oh of the net-size $s(r)$. The corresponding constant in the construction of [KP], itself based on a result of [CV], is quite large and no good bounds have been calculated for it. A recent result [PR] implies that the dual set system induced by a finite set of points and translates of an orthant in $\mathbb{R}^3$ (a generalization of $\mathcal{F}^D(P, \epsilon)$) admits an $1/r$-net of size at most $25r$ (that is efficiently constructible). Hence, for $d = 3$, problem $\mathcal{Q}$ can be efficiently approximated within a factor of 25 and the constant $c$ in (the second statement of) Theorem 4.1 is at most 25. Improving the value of this constant is an interesting open problem.

## 4.2 The Dual Problem

For a $d$-objective problem $\Pi$ with an associated GAP routine, given a parameter $k$, we want to find $k$ solution points that provide the best approximation to the Pareto curve, i.e. such that every Pareto point is $\rho^*$-covered by one of the $k$ selected points for the minimum possible ratio $\rho^* = 1 + \epsilon^*$. It was shown in [VY] that for $d = 2$ the problem is NP-hard but has a PTAS. We show below (Section 4.2.1) that for $d = 3$ any multiplicative factor for the dual problem is impossible, even for explicitly given points; we can only hope for a constant factor, and only above a certain constant.

In [VY] the dual problem was related to the asymmetric $k$-center problem, and this was used to show that (i) for any $d$, a set of $k$ points can be computed that approximates the Pareto curve with ratio $(\rho^*)^{O(\log^* k)}$, and (ii) for unbounded $d$ and explicitly given points, it is hard to do much better. Since the metric $\rho$ for the dual problem is a ratio (multiplicative coverage) versus distance (additive coverage) in the $k$-center problem, in some sense the analogue of constant factor approximation for the Dual problem is constant power. Can we achieve a constant power $(\rho^*)^c$ for all problems in MPTAS with a fixed number $d$ of objectives? We show (Section 4.2.2) that the answer is Yes for $d = 3$ and provide a conjecture that implies it for general $d$.

### 4.2.1 Lower Bound

Consider the problem $\mathcal{D}(P, k)$ (dual problem with explicitly given points): We are given *explicitly* a set $P$ of $n$ points in $\mathbb{R}_+^d$ and a positive integer $k$ and we want to compute a subset of $P$ of cardinality (at most) $k$ that $\rho$-covers $P$ with minimum ratio $\rho$. Let $\rho^* = 1 + \epsilon^*$ denote the optimal value of the ratio. Note that problems $\mathcal{Q}$ and $\mathcal{D}$ are polynomially equivalent with respect to exact optimization - as opposed to approximation. As shown in [KP], $\mathcal{Q}$ is NP-hard for $d \geq 3$; hence, for $d \geq 3$, problem $\mathcal{D}$ is also NP-hard.

By further exploiting the properties of the aforementioned reduction in [KP], we can show that problem $\mathcal{D}$ is NP-hard to approximate. Before we proceed with the formal statement and proof of this fact, it will be helpful to give some remarks regarding the notion of "approximate coverage" in the definition of the approximate Pareto set. Throughout this paper, our notion of coverage is *multiplicative*: for $\rho \geq 1$, a point $u \in \mathbb{R}_+^d$ $\rho$-covers a point $v \in \mathbb{R}_+^d$ iff $u \leq \rho \cdot v$ (coordinate-wise). Alternatively, one could define the notion of coverage additively: for $c \geq 0$, the point $u \in \mathbb{R}_+^d$ *additively* $c$-covers $v \in \mathbb{R}_+^d$ iff $u_i \leq v_i + c$ for all $i$. A notion of *additive $c$-Pareto set* can be naturally defined using the additive coverage. (Note that with the additive definition of coverage Pareto sets and approximate Pareto sets are invariant under translation of the input set, while with our multiplicative definition they are invariant under scaling.)

On the one hand, the selection of multiplicative metric is standard and more natural in the context of approximation algorithms. On the other hand, it is essential in our setting in the following sense: For a (implicitly represented) multiobjective combinatorial optimization problem, the basic existence theorem of [PY1] (i.e. the fact that there always exists an $\epsilon$-Pareto set of polynomial size) is based crucially on the multiplicative coverage. (In fact, it clearly does not hold under the additive coverage. This, of course, rules out the possibility of efficient algorithms for computing (any) approximate Pareto set in this context.) However, for the case that the set of points is given explicitly in the input (i.e. for problems $\mathcal{Q}$ and $\mathcal{D}$) the aforementioned obstacle does not occur and one can select the definition of coverage that is more appropriate for the specific application.

We will denote by $\log \mathcal{Q}$ and $\log \mathcal{D}$ the primal and dual problems respectively under additive coverage. We now try to relate the problem pairs $(\mathcal{Q}, \log \mathcal{Q})$ and $(\mathcal{D}, \log \mathcal{D})$ with respect to their approximability. To this end, we need a couple of more definitions. For two points $p, q \in \mathbb{R}_+^d$ the *ratio distance* between $p$ and $q$ is defined by: $\mathcal{RD}(p, q) = \max\{\max_i(p_i/q_i), 1\}$. (The ratio distance between $p$ and $q$ is the minimum value $\rho^* = 1 + \epsilon^*$ of the ratio $\rho$ such that $p$ $\rho$-covers $q$.) The *additive distance* between $p$ and $q$ is defined by: $\mathcal{AD}(p, q) = \max\{\max_i(p_i - q_i), 0\}$. (Analogously, the additive distance between $p$ and $q$ is the minimum value $c^*$ of the distance $c$ such that $p$ additively $c$-covers $q$.) It is easy to see that $\mathcal{AD}(\cdot, \cdot)$ is a directed pseudo-metric.

We claim that the problems $\mathcal{Q}$ and $\log \mathcal{Q}$ are in some sense "equivalent" with respect to approximability. Indeed, it is easy to see that an $r$-approximation algorithm for problem $\mathcal{Q}$ implies an $r$-approximation for problem $\log \mathcal{Q}$ and vice-versa (by taking logarithms and exponentials of the coordinates respectively). Suppose for example that there exists a factor $r$ approximation for $\log \mathcal{Q}$. We argue that it can be used as a black box to obtain an $r$-approximation for $\mathcal{Q}$. Given an instance $(P, \epsilon)$ of $\mathcal{Q}$, we construct the following instance of $\log \mathcal{Q}$: We take the set of points $P'$, where $P'$ contains a point $p'$ for every point $p \in P$ whose coordinates are the logarithms of the corresponding coordinates of $p$. We also take $c = \log(1 + \epsilon)$. That is, we ask for the smallest *additive* $c$-Pareto set of $P'$. If $p', q' \in P'$ are the images of $p, q \in P$ respectively, we have that $\mathcal{RD}(p, q) = 2^{\mathcal{AD}(p', q')}$. Hence, there exists a bijection between $\epsilon$-Pareto sets of $P$ and additive $c$-Pareto sets of $P'$, i.e. this simple transformation is an approximation factor preserving reduction of $\mathcal{Q}$ to $\log \mathcal{Q}$. There is however a subtle point regarding the bit complexity of the produced instance: the coordinates of the points in $P'$ (and the desired additive coverage $c$) may be irrational, thus not computable exactly. We argue that this is not a significant problem below.

Consider an instance $(P', c)$ of $\log \mathcal{Q}$. (The following remarks also hold for $\mathcal{Q}$ and the dual problems.) Clearly, the feasible solutions to the problem, i.e. the (additive) $c$-Pareto sets of $P'$, do not depend on the actual coordinates of the points in $P'$, but only on the additive distance between every pair of points. Hence, the only information an (exact or approximate) algorithm for $\log \mathcal{Q}$ needs to know about the input instance is the set of pairwise distances. In fact, such an algorithm does not need an explicit representation of these distances as rational numbers. It is sufficient to have a succinct representation that allows: (i) efficiently computing a succinct representation of the sum of two (or more) distances (ii) efficiently comparing any two (sums of) distances and (iii) efficiently comparing (sums of) distances with $c$. Now the aforementioned transformation produces instances $(P', c)$ of problem $\log \mathcal{Q}$ that clearly satisfy these properties (since we have an explicit representation of the starting instance $(P, \epsilon)$ of $\mathcal{Q}$ and we take logarithms). Hence, an $r$-approximation algorithm for $\log \mathcal{Q}$ can be used as a black box to obtain an $r$-approximation for $\mathcal{Q}$. Similar arguments may be used for the other direction.

For the dual problem, the choice of coverage (multiplicative versus additive) changes the objective function, which affects the approximability. Roughly speaking, a factor $r$-approximation algorithm for $\log \mathcal{D}$ is "equivalent" to a $(\rho^*)^r$-approximation algorithm for $\mathcal{D}$, where $\rho^*$ is the value of the optimal ratio for the latter problem. For example, it is easy to see (by taking logarithms as above) that a factor $r$ approximation for $\log \mathcal{D}$ implies a $(\rho^*)^r$-approximation for $\mathcal{D}$.

We have the following:

**Theorem 4.9.** *Consider the problem $\mathcal{D}(P, k)$ for $d = 3$ objectives.*
*1. It is NP-hard to approximate the minimum ratio $\rho^*$ within any polynomial multiplicative factor.*
*2. It is NP-hard to compute $k$ points that approximate the Pareto curve with ratio better than $(\rho^*)^{3/2}$.*

*Proof.* To prove both parts we take advantage of the properties in the NP-hardness reduction of [KP]. It is shown there that problem $\log \mathcal{Q}$ is NP-hard for $d = 3$ via a reduction from 3-SAT. Given an instance of 3-SAT, the reduction produces an instance $(P, c)$ of $\log \mathcal{Q}$ such that the smallest additive $c$-Pareto set of $P$ reveals whether the 3-SAT formula is satisfiable. We will not repeat the reduction here, but we will just give the properties of the construction below needed for our purposes. We prove each part separately.

1. The crucial property we need here is that the reduction in [KP] is strongly polynomial: Given an instance (formula) $\varphi$ of 3-SAT with $n$-clauses, the reduction constructs an instance of $\log \mathcal{Q}$ (or $\log \mathcal{D}$), consisting of a set $P$ of points in 3 dimensions and an additive error bound $c$ such that, if the formula $\varphi$ is satisfiable then $P$ has a (additive) $c$-cover with $g$ points (for some parameter $g$ of the construction), whereas if $\varphi$ is not satisfiable then every $c$-cover must contain at least $g + 1$ points. The construction has the property that all the points of $P$ have rational coordinates with $O(\log n)$ bits and the error bound $c \sim 1/n^2$ (to be precise, $c = 1/4n^2$). This property implies that in the (additive) dual problem $\log \mathcal{D}$ with a bound $k = g$ for the number of points in the cover, the additive "gap" in the value of the optimal covering distance between the Yes case (satisfiable 3-SAT instance $\varphi$) and the No case (non-satisfiable 3-SAT instance) is at least inverse polynomial in $n$, i.e. at least $\delta = 1/n^r$, for a (small) constant $r$: If the 3-SAT instance $\varphi$ is satisfiable, the optimal value of the covering distance for the $\log \mathcal{D}$ instance $P$ with $k = g$ is $c$; if $\varphi$ is not satisfiable, the optimal distance is at least $c' = c + \delta$. By multiplying all the coordinates of the constructed instance by a factor of $2n^{r+l}$, where $l > 0$ is a constant, and rounding to the nearest integer, we get a new instance of $\log \mathcal{D}$ where all the points have integer coordinates and the value of the additive gap between the satisfiable and the unsatisfiable case is at least $n^l$. We then exponentiate each coordinate ($x \to 2^x$). The number of bits remains polynomial in the size of the original 3-SAT instance (thus the overall reduction takes polynomial time) and the value of the *multiplicative* gap is now $2^{n^l}$.

2. To prove this part, it suffices to show that problem $\log \mathcal{D}$ does not have an approximation ratio better than $3/2$. The reduction in [KP] uses a number $g$ of gadgets. The construction has gadgets for the variables and for the clauses, which are connected by paths of flip-flop gadgets that cross using crossover gadgets. If the formula is satisfiable, then we can cover the points with additive distance $c$ with $g$ gadgets, one from each gadget. Otherwise, this is not possible. We thus select $k = g$ and ask for the "best $k$ points" and the corresponding optimal covering distance $c^*$. As previously mentioned, if the formula is satisfiable, we have $c^* = c$. Now, if the formula is not satisfiable, we argue below that the optimal covering distance is $c^* \geq 3c/2$. The proof follows directly from this.

Suppose that the 3-SAT formula is not satisfiable and we want to select the best $g$ points. First, we note that we still need one point from each gadget because otherwise all the points of a gadget must be covered by points in other gadgets that are "far away" (much further than $c$), since the gadgets are well-separated; that is, if some gadget contains no point of the solution then the covering distance is much larger than $c$. Since the formula is not satisfiable, after selecting $g$ points, at least one gadget will remain "badly covered", i.e. the point we selected must cover more points of its gadget than its $c$-neighborhood. An examination of the three types of gadgets used in the construction shows that this gives covering distance $2c$ for both the flip-flop and clause gadgets and at least $3c/2$ for the crossover gadgets. Hence, if the formula is not satisfiable, the optimal covering distance is $c^* \geq 3c/2$. ∎

### 4.2.2   Upper Bound

Consider the following *generalization* $\mathcal{Q}'(A, P, 1 + \epsilon)$ of problem $\mathcal{Q}$: Given a set of $n$ points $P \subseteq \mathbb{R}_+^d$, a subset $A \subseteq P$ and $\epsilon > 0$, compute the smallest subset $P_\epsilon^*(A) \subseteq P$ that $(1 + \epsilon)$-covers $A$. It is easy to see that for $d = 3$ the arguments of [KP, PR] for $\mathcal{Q}$ can be applied to $\mathcal{Q}'$ as well showing that it admits a constant factor approximation (see Remark 4.8). We believe that in fact for all fixed $d$ there may well be a constant factor approximation. Proving (or disproving) this for $d > 3$ seems quite challenging. The following weaker statement seems more manageable:

**Conjecture 4.10.** *For any fixed $d$, there exists a polynomial time $((1+\epsilon)^{\alpha(d)}, \beta(d))$-bicriterion approximation algorithm for $\mathcal{Q}'(A, P, 1+\epsilon)$, i.e. an algorithm that outputs an $(1+\epsilon)^{\alpha(d)}$-cover $C \subseteq P$ of $A$, satisfying $|C| \leq \beta(d) \cdot |P_\epsilon^*(A)|$, for some functions $\alpha, \beta : \mathbb{N} \to \mathbb{N}$.*

For $d = 3$, Conjecture 4.10 holds with $\alpha(3) \leq 2$, and $\beta(3) \leq 4$. This can be shown by a technical adaptation of the 3-objectives algorithm in [VY].

For general implicitly represented multiobjective problems with a polynomial $\text{GAP}_\delta$ routine, we formulate the following conjecture:

**Conjecture 4.11.** *For any fixed $d$, there exists a polynomial time generic algorithm, that outputs an $(1 + \epsilon)^{\alpha(d)}$-cover $C$, whose cardinality is $|C| \leq \beta(d) \cdot \text{OPT}_\epsilon$, for some functions $\alpha, \beta : \mathbb{N} \to \mathbb{N}$.*

The case of $d = 3$ is proved in [VY] with $\alpha(3) =$ any constant greater than 2 and $\beta(3) = 4$. Note that, by (a variant of) Lemma 4.2, Conjecture 4.10 implies Conjecture 4.11. The converse is also partially true: Conjecture 4.11 implies Conjecture 4.10, if in the statement of the latter, problem $\mathcal{Q}'$ is substituted with problem $\mathcal{Q}$.

In the following theorem, we show that a constant factor bicriterion approximation for $\mathcal{Q}'$ implies a constant power approximation for the dual problem, given the GAP routine.

**Theorem 4.12.** *Consider a (implicitly represented) $d$-objective problem in MPTAS and suppose that the minimum achievable ratio with $k$ points is $\rho^*$.*
*1. For $d = 3$ objectives we can compute $k$ points which approximate the Pareto set with ratio $O((\rho^*)^9)$, using $O((m/\delta)^d)$ $\text{GAP}_\delta$ calls, where $1/\delta = O(1/(\epsilon' - \epsilon))$.*
*2. If Conjecture 4.10 holds, then for any fixed $d$ we can compute $k$ points which approximate the Pareto set with ratio $O((\rho^*)^c)$, using $O((m/\delta)^d)$ $\text{GAP}_\delta$ calls, where $1/\delta = O(1/(\epsilon' - \epsilon))$ and $c = c(d)$.*

*Proof.* Part 1 follows from 2 since Conjecture 4.10 holds for $d = 3$. (It will follow from the proof that $c(3) \leq 9$.) To show Part 2, we exploit the relation of problem $\mathcal{D}(P, k)$ with the asymmetric $k$-center problem. As observed in [VY], the problem $\log \mathcal{D}$ is an instance of the asymmetric $k$-center problem, which we now define for the sake of completeness. In the asymmetric $k$-center problem we are given a set of $n$ vertices $V$ with distances, $\text{dist}(u, v)$ that must satisfy the triangle inequality, but may be asymmetric, i.e. $\text{dist}(u, v) \neq \text{dist}(v, u)$. We are asked to find a subset $U \subseteq V$, $|U| = k$, that minimizes $\text{dist}^* = \max_{v \in V} \min_{u \in U} \text{dist}(u, v)$. (Note that $\log \mathcal{D}(P, k)$ is an instance of this problem, where there exists a bijection between vertices of $V$ and points of $P$ and the distance between points (vertices) $p, q \in P$ is defined as $d(p, q) = \mathcal{AD}(p, q)$.)

We claim that, if problem $\mathcal{Q}'(A, P, 1 + \epsilon)$ admits a $((1 + \epsilon)^{\alpha(d)}, \beta(d))$-bicriterion approximation, then problem $\mathcal{D}(P, k)$ admits a $(\rho^*)^{c(d)}$ approximation for some function $c$ (that depends on $\alpha$ and $\beta$). This is implied by the aforementioned reduction and the following more general fact: If we have an instance of the asymmetric $k$-center problem (problem $\log \mathcal{D}(P, k)$ in our setting) such that a certain collection of associated set cover subproblems (which are instances of problem $\log \mathcal{Q}'(A, P, 1 + \epsilon)$ here) admits a constant factor bicriterion approximation (an algorithm that blows up both criteria by a constant factor), then this instance admits a constant factor *unicriterion* approximation (an algorithm that outputs a set of no more than $k$ centers). This implication is not stated in [PV, Ar1], but is implicit in their work. One way to prove it is to apply Lemma 5 of [PV] in a recursive manner. We will describe an alternative method [Ar2] that yields better constants. We prove this implication, appropriately translated to our setting, in Lemma 4.14.

For a general multiobjective problem where the solution points are not given explicitly, we impose a geometric $\sqrt{1 + \delta}$ grid for a suitable $\delta$, call $\text{GAP}_\delta$ at the grid points, and then apply the above algorithm to the set of points returned. Then the set of $k$ points computed by the algorithm provide a $(1 + \epsilon')^{c(d)}$-cover of the Pareto curve, where $1 + \epsilon' = (1 + \epsilon)(1 + \delta)^2$. ∎

*Remark* 4.13. Even though the $O(\log^* k)$-approximation ratio is best possible for the (general) asymmetric $k$-center problem [CG+], the corresponding hardness result does not apply for $\log \mathcal{D}$ as long as the dimension $d$ is fixed.

Let $H(\alpha)$ denote the harmonic number extended to fractional arguments by linear interpolation (i.e. $H(\alpha) = \sum_{i=1}^{\lfloor \alpha \rfloor} 1/i + (\alpha - \lfloor \alpha \rfloor)/\lceil \alpha \rceil$). For a function $g$, let $g^{(i)}$ denote the function iterated $i$ times. Finally, for $b > 1$ define $H_b^*(\alpha) = \min\{i : H^{(i)}(\alpha) \leq b\}$. The following lemma completes the proof of Theorem 4.12.

**Lemma 4.14.** *Suppose that there exists an $((1 + \epsilon)^\alpha, \beta)$-bicriterion approximation for $\mathcal{Q}'(A, P, 1 + \epsilon)$. Then, problem $\mathcal{D}(P, k)$ admits a $(\rho^*)^c$ approximation, where $c = H_{4/3}^*(\beta) + \alpha + 4$. In particular, for $\alpha = 2$ and $\beta = 4$, we can get $c = 9$.*

*Proof.* The desired result can be shown by a careful application of the techniques introduced in [PV, Ar1]. We describe an algorithm – that we denote $\mathcal{D}(P, k)$, as the corresponding problem – which, given a $(\rho^\alpha, \beta)$-bicriterion approximation algorithm, denoted $\mathcal{B}(A, P, \rho)$, for problem $\mathcal{Q}'(A, P, \rho)$ as a black box, computes a set $Q \subseteq P$ of (at most) $k$ points that $(\rho^*)^c$-cover the set $P$, where $\rho^*$ is the minimum ratio achievable with $k$ points. We will denote by $\mathcal{B}(A, P, \rho)$ the set of points output by the algorithm $\mathcal{B}$ on input $(A, P, \rho)$.

We first note the simple (and well-known) fact that it is no loss of generality to assume that the algorithm $\mathcal{D}(P, k)$ "knows" the optimal ratio $\rho^*$; this is because $\rho^*$ will be one of the $O(|P|^2)$ pairwise ratio distances, hence we can try the algorithm for all of them and pick the best solution (or do an appropriate binary search, see e.g. [Ar1]).

To describe the algorithm, we appropriately translate the notions from [PV, Ar1] to the current setting. For the sake of completeness, we also provide a mostly complete proof of correctness.

We begin with a basic definition.

**Definition 4.15.** For a point $q \in P$ and a parameter $\rho > 1$, we denote $\Gamma^+(q, \rho) = \{p \in P \mid q \leq \rho \cdot p\}$ the set of points in $P$ $\rho$-covered by $q$ and $\Gamma^-(q, \rho) = \{p \in P \mid p \leq \rho \cdot q\}$ the set of points in $P$ that $\rho$-cover $q$. We naturally extend this notation to sets $S \subseteq P$: $\Gamma^\pm(S, \rho) = \{p \in P \mid p \in \Gamma^\pm(s, \rho) \text{ for some } s \in S\}$. We say that the point $q \in P$ is a *$\rho$-center capturing vertex* (denoted $\rho$-CCV) if it satisfies $\Gamma^-(q, \rho) \subseteq \Gamma^+(q, \rho)$.

Consider an instance of the problem $\mathcal{D}(P, k)$ as defined above. Suppose that $\rho \geq \rho^*$. In this case, if the point $q$ is a $\rho$-CCV, it $\rho$-covers at least one point of the optimal solution – in particular, the point $q^*$ that $\rho^*$-covers $q$. Indeed, $q^* \in \Gamma^-(q, \rho^*) \subseteq \Gamma^-(q, \rho) \subseteq \Gamma^+(q, \rho)$. Hence, $q$ $\rho^2$-covers every point in $P$ $\rho$-covered by the point $q^*$. This simple property is crucial for the algorithm.

The algorithm in [PV] has two phases. In the first phase, roughly, it preprocesses the input set by iteratively finding CCV's and in the second phase it uses a recursive set cover procedure to cover the points not covered in the first stage. (The algorithm in [Ar1] replaces the second phase by an LP-based method.)

The algorithm $\mathcal{D}(P, k)$ works in three phases. The first phase is identical to the first phase in [PV, Ar1]: We preprocess the input set $P$ by iteratively finding $\rho^*$-CCVs. In the second phase, $\mathcal{D}$ calls the bicriterion approximation algorithm $\mathcal{B}$ (with appropriately selected values of its parameters) to cover the subset of $P$ that is not covered in the first phase. The remaining phase involves a careful application of the recursive greedy set cover procedure of [PV] followed by an application of the greedy set cover algorithm. To show correctness of the last step, we use the structural lemma of [Ar1] (itself a variant of a similar lemma in [PV], albeit with improved constants).

The algorithm is presented in detail below:

We now proceed with an intuitive explanation of the different steps in tandem with a proof of correctness. We explain first what happens during the first phase. We have as input the set $P$, the parameter $k$ and the optimal ratio $\rho^*$. (Recall that the algorithm can "guess" the optimal ratio.) We iteratively select $\rho^*$-CCV's as follows: For each $\rho^*$-CCV we find, we remove from the "active" set $A$ (initialized to $P$) all the points $(\rho^*)^2$-covered by it, until no more CCV's exist in $A$. Let $C$ be the set of CCV's thus discovered ($|C| \leq k$) and $A = P \setminus \Gamma^+(C, (\rho^*)^2)$ be the set of points in $P$ not $(\rho^*)^2$-covered by any point in $C$. At this point, we note the following simple fact:

**Fact 4.16.** *The set $A := P \setminus \Gamma^+(C, (\rho^*)^2)$ can be $\rho^*$-covered by $k' = k - |C|$ points in $P \setminus \Gamma^+(C, \rho^*)$.*

If $|C| = k$ ($k' = 0$, $A = \emptyset$), we have selected a set of $k$ points that $(\rho^*)^2$-cover the set $P$ and we can just terminate the algorithm. Otherwise, we proceed with the next phase. In the second phase, we call the algorithm $\mathcal{B}$ to $\rho^*$-cover the set $A$. By Fact 4.16, there exists a $\rho^*$-cover of $A$ with $k'$ points. Moreover, it is clear that such a cover lies in $P \setminus \Gamma^+(C, \rho^*)$. Hence, we get a set $S_0 \subseteq P \setminus \Gamma^+(C, \rho^*)$ of cardinality $|S_0| \leq \beta \cdot k'$ that $(\rho^*)^\alpha$-covers $A$. To motivate the next step, we note the following immediate implication of Fact 4.16:

**Algorithm** $\mathcal{D}(P, k)$

(The optimal radius $\rho^*$ is known to the algorithm.)

(***Phase 1***)

$A = P$; $k' = k$; $C = \emptyset$;

**While** there exists a $\rho^*$-CCV $q \in A$ and $k' > 0$ **do**

$\{ C = C \cup \{q\}$;

    $A = A \setminus \Gamma^+(q, (\rho^*)^2)$;

    $k' = k' - 1$; $\}$

(***Phase 2*** )

$S_0 = \mathcal{B}(A, P \setminus \Gamma^+(C, \rho^*), \rho^*)$;

(***Phase 3***)

$\widehat{S}_0 = S_0 \setminus \Gamma^+(C, (\rho^*)^2)$;

$S_1 = $ Rec-Cover $(\widehat{S}_0, A, P, \rho^*, k')$;

$\widehat{S}_1 = S_1 \setminus \Gamma^+(C, (\rho^*)^4)$;

$S_2 = $ Greedy-Set-Cover $(\widehat{S}_1, P, (\rho^*)^3)$;

**Return** $Q := C \cup S_2$.

**Routine** Rec-Cover (*Input:* $S, A, P, \rho, l$)

(There exist $l$ vertices in $P$ that $\rho$-cover $S$,

  where $S \subseteq A \subseteq P$.)

$S^0 = S$; $i = 0$;

**While** $|S^i| > 4l/3$ **do**

$\{$

  Run Greedy Set Cover to $\rho$-cover $S^i$ using points
of $P$ and let $\widetilde{S}^{i+1} \subseteq P$ be the produced set.

  $S^{i+1} = \widetilde{S}^{i+1} \cap A$;

  $i = i + 1$;

$\}$

**Return** $S^i$.

**Fact 4.17.** *Let $S \subseteq A$. Then $S$ can be $\rho^*$-covered by $k'$ points in $P \setminus \Gamma^+(C, \rho^*)$.*

We also recall the following well-known fact [Chv, Joh, Lov] about the performance guarantee of the greedy set cover algorithm:

**Fact 4.18.** *For a set system $(U, \mathcal{R})$ suppose that there exists a set cover of cardinality $p$. Then the greedy algorithm outputs a cover of size at most $p \cdot H(|U|/p)$.*

At this point, we apply the recursive greedy set cover procedure from [PV] to cover $\widehat{S}_0 = S_0 \cap A$ using points from $A$. (The points in $S_0 \setminus \widehat{S}_0$ are $(\rho^*)^2$-covered by $C$.) Note that in each round of the recursive cover, we attempt to cover only those points from the last round that do not lie in $\Gamma^+(C, (\rho^*)^2)$, since $C$ will cover those ones. We thus get a set $S_1 \subseteq P$ of cardinality $|S_1| \leq 4k'/3$ with the property that $S_1$ covers $S_0 \setminus \Gamma^+(C, (\rho^*)^{1+H^*_{4/3}(\beta)})$ with ratio $(\rho^*)^{H^*_{4/3}(\beta)}$. The latter statement can be shown by induction, using Fact 4.17 as an invariant. Since this essentially appears in [PV, Ar1] (see e.g. Lemma 13 in [Ar1]), we do not repeat it here. To motivate the next step, we need the following combinatorial lemma from [Ar1]:

**Lemma 4.19** (Theorem 17 in [Ar1], rephrased). *Let $C \subseteq P$ and $A = P \setminus \Gamma^+(C, (\rho^*)^2)$. Suppose $A$ has no $\rho^*$-CCV's and that there exist $k'$ centers (points in $P$) that $\rho^*$-cover $A$. Then there exists a set of $2k'/3$ centers in $P \setminus \Gamma^+(C, \rho^*)$ that $(\rho^*)^3$-covers $A' = P \setminus \Gamma^+(C, (\rho^*)^4)$.*

As a final step of the algorithm, we apply the greedy set cover algorithm – that may be viewed as one iteration of the recursive procedure – with parameter $(\rho^*)^3$ to cover $\widehat{S}_1 = S_1 \setminus \Gamma^+(C, (\rho^*)^4)$ using points from $P$ (so that the optimum has cardinality at most $2k'/3$, according to Lemma 4.19). (Note that the points in $S_1 \setminus \widehat{S}_1$ are $(\rho^*)^4$-covered by $C$.) We thus get a set $S_2 \subseteq P$ of cardinality at most $(2k'/3) \cdot H((4k'/3)/(2k'/3)) = (2k'/3) \cdot H(2) = k'$ with the property that $S_2$ covers $\widehat{S}_1$ within $(\rho^*)^3$. We output the set $Q := C \cup S_2$; this set has cardinality at most $k$ and it remains to argue that it covers $P$ with ratio $(\rho^*)^{4+\alpha+H^*_{4/3}(\beta)}$. Indeed, every point $p \in P$ falls in one of the following categories:

- The point $p$ is $(\rho^*)^2$-covered by a point in $C$, i.e. $p \in \Gamma^+(C, (\rho^*)^2)$. (Note that if this is *not* the case, i.e. if $p \in A$, then it is $(\rho^*)^\alpha$-covered by $S_0$.)

- The point $p$ is $(\rho^*)^\alpha$-covered by a point $p_0 \in S_0$ that is *not* $(\rho^*)^{H^*_{4/3}(\beta)}$ - covered by $S_1$. In this case, $p_0 \in \Gamma^+(C, (\rho^*)^{1+H^*_{4/3}(\beta)})$, so $C$ covers $p$ within ratio $(\rho^*)^{1+H^*_{4/3}(\beta)+\alpha}$.

- The point $p$ is $(\rho^*)^\alpha$-covered by a point $p_0 \in S_0$ that is $(\rho^*)^{H^*_{4/3}(\beta)}$ - covered by a point $p_1 \in S_1$ that is *not* $(\rho^*)^3$-covered by $S_2$. In this case, $p_1 \in \Gamma^+(C, (\rho^*)^4)$, so $C$ covers $p$ within ratio $(\rho^*)^{4+H^*_{4/3}(\beta)+\alpha}$.

- The point $p$ is $(\rho^*)^\alpha$-covered by a point $p_0 \in S_0$ that is $(\rho^*)^{H^*_{4/3}(\beta)}$ - covered by a point $p_1 \in S_1$ that is in turn $(\rho^*)^3$-covered by $p_2 \in S_2$. In this case, the point $p_2$ covers $p$ within ratio $(\rho^*)^{3+H^*_{4/3}(\beta)+\alpha}$.

Hence the overall covering ratio is $(\rho^*)^{4+H^*_{4/3}(\beta)+\alpha}$, which completes the proof. ∎

*Remark* 4.20. We note here that the recursive set cover procedure (used in the above lemma) was useful merely to improve the constants in the reduction. One can alternatively prove a (quantitatively inferior) version of the lemma by the following two-phase algorithm: In the first phase, preprocess the input set $P$ by iteratively finding $\rho$-CCVs for appropriately chosen values of the parameter $\rho$. In the second phase, call the algorithm $\mathcal{B}$ to "cover" the subset of $P$ that is not covered in the first phase. The analysis of this alternative algorithm is based on repeated applications of Lemma 4.19.

*Remark* 4.21. We should remark that the algorithms of this section are less satisfactory that the bi-objective algorithm of the previous section (and the 2-d and 3-d algorithms of [VY]) in several respects. One weakness is that the constants $c$ obtained (for $d = 3$) are quite large: in the case of Theorem 4.1, the best constant $c$ we can get follows from the net construction of [PR] (and is about 25). In the case of Theorem 4.12 there is still a large gap between the upper bound (of 9) and the lower bound (of $3/2$) in the exponent.

A second weakness of the algorithms is that they start by applying the general method of [PY1] calling the GAP routine on a grid, and thus incur always the worst-case time complexity even if there is a very small $\epsilon$-Pareto set. Thus, we view our algorithms in this section mainly as theoretical proofs of principle, i.e. that certain (constant) approximations can be computed in polynomial time, but it would be very desirable and important to improve both the constants and the time.

## 5 Conclusion

We investigated the problem of computing a minimum set of solutions for a multiobjective optimization problem that represents approximately the whole Pareto curve within a desired accuracy $\epsilon$. We developed tight approximation algorithms for the bi-objective shortest path problem, spanning tree, and a host of other bi-objective problems. Our algorithms compute efficiently an approximate Pareto set that contains at most twice as many solutions as the minimum one; furthermore improving on the factor 2 for these specific problems is NP-Hard. The algorithm works in general for all bi-objective problems for which we have a routine for the Restricted problem of approximating one objective subject to a (hard) bound on the other. The algorithm calls this Restricted routine and a dual one as black boxes and makes quite effective use of them: for every instance, the number of calls is linear (at most 4 times) in the number of points in the optimal solution for that instance.

We presented also results for three and more objectives, both for the problem of computing an optimal $\epsilon$-Pareto set and for the dual problem of selecting a specified number $k$ of points that provide the best approximation of the full Pareto curve. As we indicated at the end of the last section, there is still a lot of room for improvement both in the time complexity and the constants of the approximations achieved. We would like especially to resolve Conjecture 4.11, hopefully positively. It would be great to have a general efficient method for any (small) fixed number $d$ of objectives that computes for every instance a succinct approximate Pareto set with small constant loss in accuracy and in the number of points, and do it in time proportional to the number of computed points, i.e., the optimal approximate Pareto set for the instance in hand.

**Acknowledgements.** We would like to thank Aaron Archer for useful discussions related to the asymmetric $k$-center problem.

# References

[ABK1]  E. Angel, E. Bampis, A. Kononov. An FPTAS for Approximating the Unrelated Parallel Machines Scheduling Problem with Costs. *In Proc. ESA*, pp. 194–205, 2001.

[ABK2]  E. Angel, E. Bampis, A. Kononov. On the approximate trade-off for bicriteria batching and parallel machine scheduling problems. *Theoretical Computer Science*, 306(1-3), pp. 319–338 (2003).

[ABV1]  H. Aissi, C. Bazgan, D. Vanderpooten. Approximation Complexity of Min-Max (Regret) Versions of Shortest Path, Spanning Tree and Knapsack. In *Proc. ESA*, pp. 862–873, 2005.

[ABV2]  H. Aissi, C. Bazgan, D. Vanderpooten. Complexity of Min-Max (Regret) Versions of Cut problems. In *Proc. ISAAC*, pp. 789–798, 2005.

[AN+]  H. Ackermann, A. Newman, H. Röglin, and B. Vöcking. Decision Making Based on Approximate and Smoothed Pareto Curves. *Theoretical Computer Science*, 378(3), p. 253–270 (2007).

[Ar1]  A. F. Archer. Two $O(\log^* k)$-approximation algorithms for the asymmetric $k$-center problem. In *Proc. IPCO*, pp. 1–14, 2001.

[Ar2]  A. F. Archer. Personal communication, 2007.

[As]  P. Assouad. Densité et Dimension. *Ann. Institut Fourier, Grenoble*, 3:232–282, 1983.

[AZ]  A. Armon, U. Zwick. Multicriteria Global Minimum Cuts. *Algorithmica*, 46, pp. 15–26, 2006.

[BG]  H. Brönnimann, M.T. Goodrich. Almost Optimal Set Covers in Finite VC-Dimension. *Discrete and Computational Geometry*, 14(4): 463–479 (1995).

[Chv]  V. Chvátal. A greedy heuristic for the set covering problem. *Math. Oper. Res.*, 4: 233–235 (1979).

[CG+]  J. Chuzhoy, S. Guha, E. Halperin, S. Khanna, G. Kortsartz, R. Krauthgamer, J. Naor. Asymmetric $k$-center is $\log^* n$-hard to Approximate. *J. ACM*, 52(4): 538–551 (2005).

[CJK]  T. C. E. Cheng, A. Janiak, M. Y. Kovalyov. Bicriterion Single Machine Scheduling with Resource Dependent Processing Times. *SIAM J. Optimization*, 8(2), pp. 617–630, 1998.

[CX]  G. Chen, G. Xue. A PTAS for weight constrained Steiner trees in series-parallel graphs. *Theoretical Computer Science*, 1-3(304), pp. 237–247, 2003.

[CX2]  G. Chen, G. Xue. $k$-pair delay constrained minimum cost routing in undirected networks. In *Proc. SODA*, pp. 230–231, 2001.

[Cli]  J. Climacao, Ed. *Multicriteria Analysis.* Springer–Verlag, 1997.

[CV]      K. L. Clarkson, K. Varadarajan. Improved approximation algorithms for geometric set cover. *Discrete & Computational Geometry*, 37(1):43-58, 2007.

[DJSS]   J. Dongarra, E. Jeannot, E. Saule, Z. Shi. Bi-objective Scheduling Algorithms for Optimizing Makespan and Reliability on Heterogeneous Systems. In *Proc. SPAA*, pp. 280–288, 2007.

[Ehr]     M. Ehrgott. *Multicriteria optimization*, 2nd edition, Springer–Verlag, 2005.

[EG]      M. Ehrgott, X. Gandibleux. An annotated bibliography of multiobjective combinatorial optimization problems. *OR Spectrum* 42, pp. 425–460, 2000.

[ERS]    G. Even, D. Rawitz, S. Shahar. Hitting sets when the VC-dimension is small. *Information Processing Letters*, 95(2):358–362, 2005.

[ESZ]    F. Ergun, R. Sinha, L. Zhang. An improved FPTAS for Restricted Shortest Path. *Information Processing Letters*, 83(5):237–239, 2002.

[EV]      E. Erkut, V. Verter. Modeling of transport risk for hazardous materials. *Operations Research*, 46, pp. 625–642, 1998.

[Fei]     U. Feige. A threshold of $\ln n$ for approximating set cover. *JACM*, 45(4), pp. 634–652, 1998.

[FGE]    J. Figueira, S. Greco, M. Ehrgott, eds. *Multiple Criteria Decision Analysis: State of the Art Surveys*, Springer, 2005.

[GJ]      M. R. Garey, D. S. Johnson. *Computers and Intractability*. W. H. Freeman, 1979.

[GR]      M.X. Goemans, R. Ravi. The Constrained Minimum Spanning Tree Problem. In *Proc. SWAT*, 1996, pp. 66–75.

[GR+]    A. Goel, K. G. Ramakrishnan, D. Kataria, and D. Logothetis. Efficient computation of delay–sensitive routes from one source to all destinations. In *Proc. IEEE INFOCOM*, 2001.

[Han]    P. Hansen. Bicriterion Path Problems. In *Proc. 3rd Conf. Multiple Criteria Decision Making Theory and Application*, pp. 109–127, Springer Verlag LNEMS 177, 1979.

[Has]    R. Hassin. Approximation schemes for the restricted shortest path problem. *Mathematics of Operations Research*, 17(1), pp. 36–42, 1992.

[HL]     R. Hassin, A. Levin. An efficient polynomial time approximation scheme for the constrained minimum spanning tree problem. *SIAM J. Comput.*, 33(2): 261–268 (2004).

[HW]     D. Haussler, E. Welzl. Epsilon-nets and simplex range queries. *Discrete Computational Geometry*, 2:127–151, 1987.

[Joh]    D. S. Johnson. Approximation algorithms for combinatorial problems. *J. Comput. System Sci.*, 9: 256–278 (1978).

[KP]     V. Koltun, C.H. Papadimitirou. Approximately dominating representatives. *Theoretical Computer Science*, 371, pp. 148–154, 2007.

[KPW]   J. Komlos, J. Pach, W. Woeginger. Almost tight bounds for Epsilon-Nets. *Discrete and Computational Geometry*, 7, pp. 10–15, 1992.

[Lov]   L. Lovasz. On the ratio of optimal integral and fractional covers. *Discrete Math.*, 13: 383–390 (1975).

[LR]    D. H. Lorenz, D. Raz. A simple efficient approximation scheme for the restricted shortest path problem. *Operations Research Letters*, 28(5), pp. 213–219, 2001.

[LY]    C. Lund, M. Yannakakis. On the hardness of approximating minimization problems. *JACM*, 41(5), pp. 960–981, 1994.

[Mit]   K. M. Miettinen. *Nonlinear Multiobjective Optimization*, Kluwer, 1999.

[PR]    E. Pyrga, S. Ray. New Existence Proofs for $\epsilon$-Nets. In *Proc. SoCG*, 2008.

[PV]    R. Panigrahy, S. Vishwanathan. An $O(\log^* n)$ approximation algorithm for the asymmetric $p$-center problem. *J. of Algorithms*, 27(2), pp. 259–268, 1998.

[PW]    J. Pach, W. Woeginger. Some new bounds for Epsilon-Nets. In *Proc. 6th ACM Symposium on Computational Geometry*, pages 10–15, 1990.

[PY1]   C.H. Papadimitriou, M. Yannakakis. On the Approximability of Trade-offs and Optimal Access of Web Sources. In *Proc. FOCS*, pp. 86–92, 2000.

[PY2]   C.H. Papadimitriou, M. Yannakakis. Multiobjective Query Optimization. In *Proc PODS*, pp. 52–59, 2001.

[TZ]    G. Tsaggouris, C.D. Zaroliagis. Multiobjective Optimization: Improved FPTAS for Shortest Paths and Non-linear Objectives with Applications. In *Proc. ISAAC*, pp. 389–398, 2006.

[VV]    P. Van Mieghen, L. Vandenberghe. Trade-off Curves for QoS Routing. In *Proc. INFOCOM*, 2006.

[VC]    V. N. Vapnik, A. Ya. Chervonenkis. On the uniform convergence of relative frequencies to their probabilities. *Theory Probab. Appl.*, 16(2): 264-280, 1971.

[VY]    S. Vassilvitskii, M. Yannakakis. Efficiently computing succinct trade-off curves. *Theoretical Computer Science* 348, pp. 334–356, 2005.

[Wa]    A. Warburton. Approximation of Pareto Optima in Multiple-Objective Shortest Path Problems. *Operations Research*, 35, pp. 70–79, 1987.